\documentclass[%
 reprint,
 amsmath,amssymb,
 aps,
]{revtex4-2}
\usepackage{CJK}
\usepackage{graphicx}
\usepackage{dcolumn}
\usepackage{bm}

\begin{document}
\begin{CJK*}{GB}{}
\preprint{APS/123-QED}

\title{Experimental investigation of the exit dynamics of a horizontal circular
	 cylinder out of water and silicone oil}

\author{Intesaaf Ashraf$^1$, Lionel Vincent$^2$, Romain Falla$^3$, Vincent E. Terrapon$^3$, Benoit Scheid$^2$, and St\'ephane Dorbolo$^1$}
 
\affiliation{$^1$ PtYX, D\'epartement de Physique, Universit\'e de Li\`ege (ULi\`ege), 4000 Li\`ege, BE \\
$^2$ Transfers, Interfaces and Processes (TIPs), Universit\'e Libre de Bruxelles (ULB), 1050 Brussels, BE \\
$^3$ Aerospace and Mechanical Engineering, Universit\'e de Li\`ege (ULi\`ege), Li\`ege, BE
}

\date{\today}

\begin{abstract}
Experimental investigations of the exit dynamics of a horizontal cylindrical object were performed in water and silicone oil (50 cSt). The fully immersed cylinder was initially at rest in a still fluid tank before being pushed (or pulled according to the measurement procedure) upwards at a constant velocity. The cylinder geometry was chosen to approach two-dimensional conditions in the transverse direction and, thus, allow comparison with two-dimensional numerical simulations. Firstly, we demonstrate that these conditions are better satisfied for a large aspect ratio cylinder equipped with vertical end-plates. Secondly, the influence of the initial depth on the liquid entrained and the wake generated by the cylinder is discussed.  It is shown that the wake destabilizes in the form of symmetry breaking in the case of water, while it remains symmetrical in the case of oil even for long travel in the bath. The deformation of the interface is found to be independent of the starting depth when the starting depth is larger than 6 times the cylinder diameter. In the present case, this criterion reflects also the finite acceleration of the cylinder to reach the determined constant exit velocity. Measurements in a range of exit speeds between 0.1 and 1 m/s indicate that the thickness of the liquid above the cylinder when the cylinder starts crossing the interface, increases with the speed according to a logarithmic law of the Froude number. During the subsequent drainage, the evolution of the coated liquid thickness is found to first decrease exponentially with time just after the crossing of the interface. At later times, a change of regime occurs and the drainage follows the inverse of the square root of time irrespective of the crossing speed. Finally, the force necessary to maintain a constant exit speed during the motion of the cylinder inside and outside the bath is analyzed. This global measurement of the entrained liquid confirms the square root scaling of the thinning with time during the drainage process.
\end{abstract}

\maketitle
\end{CJK*}

\section{Introduction}\label{intro}
The physics of a body entering or exiting a liquid bath has been the subject of numerous studies, mostly driven in recent years by space (e.g., the splashdown of spacecraft), military (e.g., missile launching by submarines), and industrial (e.g., coating) applications. Water entry and exit problems are also commonly found in nature, for instance, animals jumping out of the water \cite{chang2019jumping} or animals diving in the water \cite{dive}. In both cases, accurate experimental measurements, theoretical analysis, and numerical simulations are challenging because of the complexity of the phenomena to take into account, such as interfacial physics and fluid-structure interactions.

While research on water entry has primarily focused on optimum design, shape, and entry velocity \cite{challa2014, challa2010, mohtat2015, yang2012,nair2018water, projectile}, the water entrainment after the exit is of particular interest.  Indeed, the exit problem is closely related to the drainage process of the entrained liquid. Depending on the applications, fast drainage might be required (missiles, fishing birds), or, on the other hand, slow and uniform drainage might be more beneficial.  For instance, when an object has to be coated with paint, the simplest way to proceed is to plunge the object into a large bath and pull it out of the bath. The speed of pulling, the viscosity, and the surface tension of the fluid, as well as the contact angle, have to be taken into account for the coating of the surface parallel to the pulling direction.  

This problem of the extraction of a solid from a liquid was addressed in the work by Landau, Levich and Derjaguin (LLD) in the case of a vertical plate that is extracted from viscous liquids and/or with small velocities \cite{LL, LL2}. The configuration is rather generic and is found in numerous industrial processes involving the dip-coating of a surface to transform its property (coloring, anti-scratching, anti-wetting, anti-oxidizing, etc...). Studies have shown the richness of the LLD problem by considering the effects on the liquid entrainment of complex fluids \cite{deryck}, solid micro-textures \cite{seiwert}, and solid elasticity \cite{dixit}. These researches have improved the reliability and the control of the dip-coating process. Some works have also considered the influence of the inclination, either of the substrate \cite{benilov}, of the pulling direction \cite{weinstein}, or even along the bottom face of the object \cite{jambon, eghbali}. The subsequent liquid film drainage along flat or curved surfaces is another field that has extensively been studied since Reynolds' seminal work \cite{Reynolds} and is still very active \cite{drainage, coons}. However, the LLD problem refers to objects that are partially immersed in their initial position. The total amount of liquid entrained by lifting an object as simple as a sphere out of a bath remains an open question \cite{bhamla}. 

The manner to extract the object from the bath is a key to determining the drainage dynamics. Concerning the water exit problem, the extraction can be categorized according to different parameters. The body motion can be forced through an imposed displacement \cite{wu2017experimental,liju2001} or an external force \cite{moshari2014}, driven by buoyancy \cite{truscott2016} (e.g., when playing with a balloon in a swimming pool \cite{guyon})  or a combination of an initial forced motion followed by an inertial phase \cite{takamure2021effect}. Regarding the shape of the body, most works have focused on rather canonical shapes, including spheres \cite{wu2017experimental,truscott2016,takamure2021effect,intesaaf1}, vertical cylinders \cite{liju2001,chu2010}, horizontal cylinders \cite{Havelock1936,greenhow1983nonlinear,greenhow1997,telste1987}, prolate ellipses \cite{ni2015}, or even square cylinder \cite{intesaaf2}.

In the present work, we focus on the extraction of a horizontal cylinder out of a bath of liquid with a constant velocity.  The motivation for studying the horizontal cylinder exit is to restrict the flow around the object to mimic and approximate two-dimensional conditions for the flow around the section of the object. We chose to study constant speed conditions since they are encountered in numerous situations and since the speed during the crossing of the interface can be approximated to be constant. The water exit of a horizontal cylinder has already been addressed in some previous works.  

Greenhow {\it et al.} \cite{greenhow1983nonlinear} performed inspiring experimental works. A few years later, the same group \cite{greenhow1997} published two-dimensional numerical simulations of the free-surface deformation of an initially calm water surface.  In this paper, the experimental data set can be found for a ``short'' cylinder (the length being two or three times the diameter of the cylinder). 

A more extensive work was proposed by Haohao {\it et al.} \cite{haohao2019numerical}, including the case of the sphere. Both numerical simulations and some experimental results were presented. A cylinder with an aspect ratio (length divided by radius) of 4 was equipped with vertical end-plates to keep the velocity field parallel to the direction of motion. The measurements were compared to a model CIP (Constrained Interpolation Profile) based on a work by Miao {\it et al.} \cite{miao1989} with good agreement for a reduced set of imposed conditions.  Additionally, simulations were carried out at four different Froude numbers (defined as the square of the object's speed divided by the gravity times the radius of the object). The free-surface elevation when the cylinder starts crossing the interface was found to be strongly dependent on the Froude number for values below $4.12$. However, for large speeds (Froude number between $4.12$ and $8.24$), the time dependence of the surface elevation was nearly the same. Finally, these authors reported that waterfall breaking becomes more intense as the Froude number increases due to the vortex generated in the wake. On the other hand, the waterfall reduces to a ligament at low speeds. Note that, beyond the waterfall, the liquid that flows around the cylinder may also destabilize into a ligament. In this case, the coating thickness was reported numerically in a recent paper \cite{wei2024}.

\begin{figure*}
		\centering
	
		\includegraphics[width=0.7\textwidth,angle=-90]{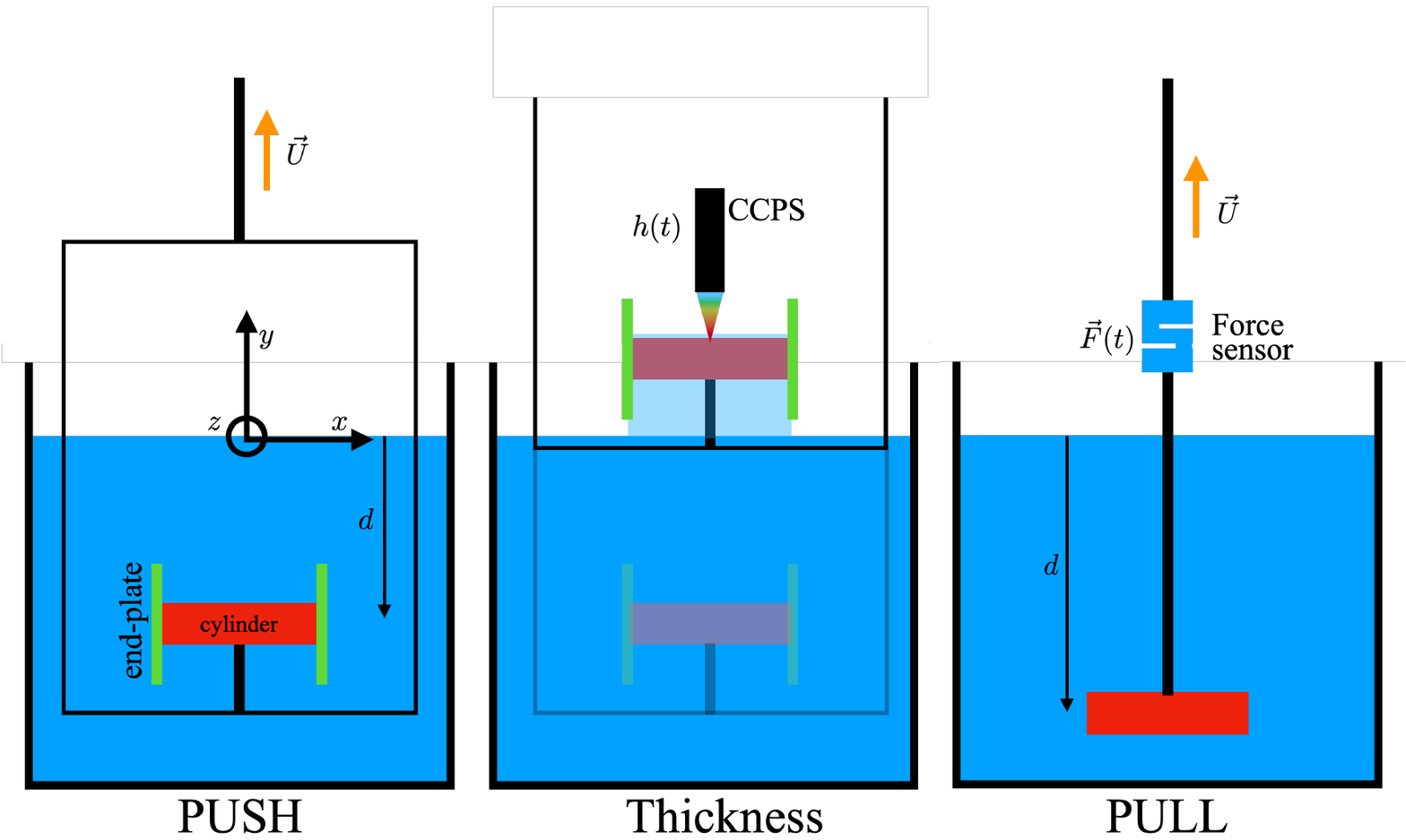}
			\includegraphics[width=5cm]{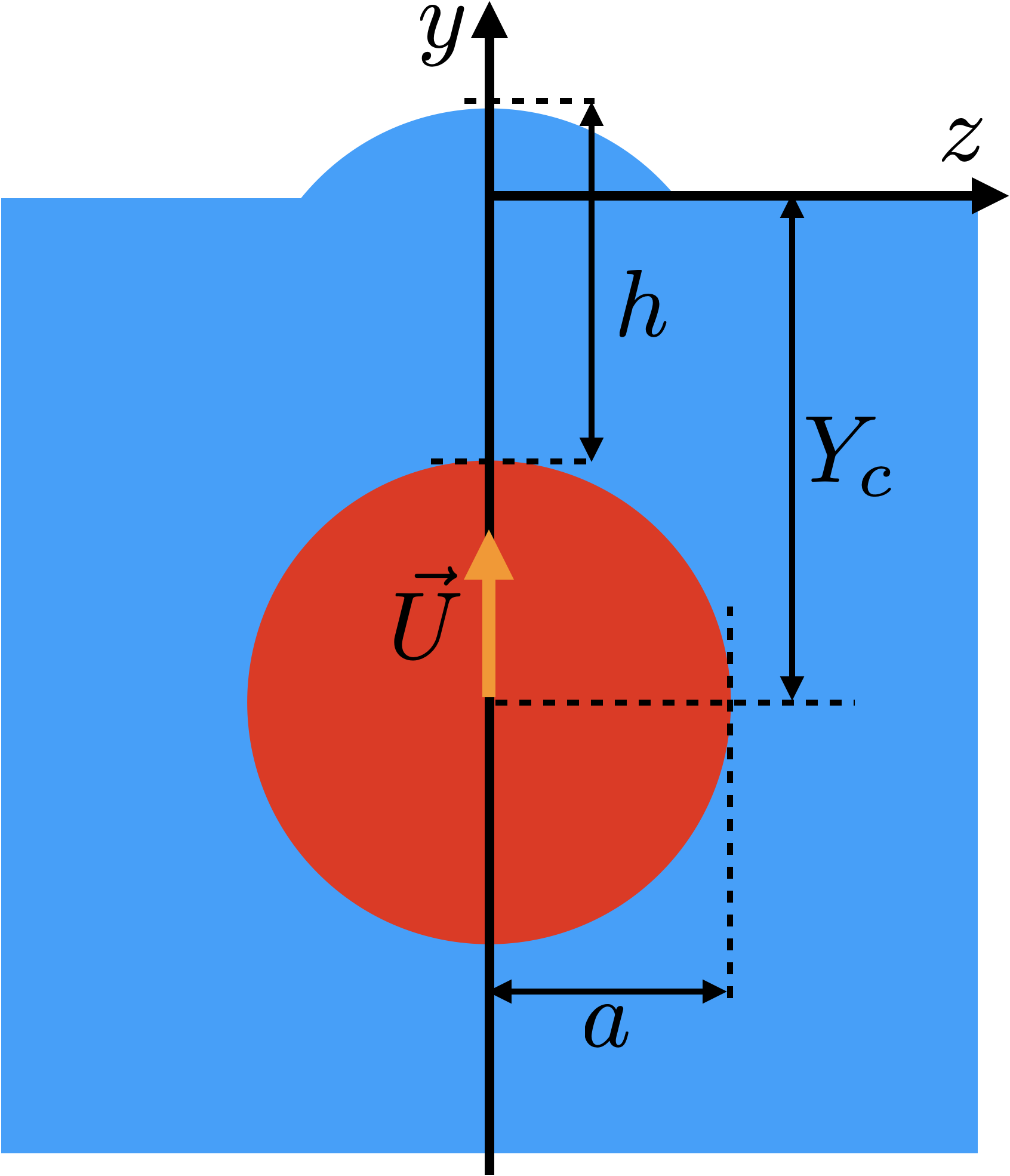}
		
		\caption{(top) Schematic representations of the experimental configurations considered in this work. Push mode: the cylinder was attached by the bottom and pushed upwards via a light rigid frame. End plates were added for some experiments to improve 2D conditions. Pull mode: the cylinder was attached by the top in order to limit the impact of the frame in the force measurements. Central panel: the cylinder can be stopped at a defined position to measure the oil film thickness evolution with time using a Chromatic Confocal Point Sensor (CCPS). (bottom) Sketch of the different parameters used in the work.}
		\label{fig1}
	\end{figure*}
	
Moshari {\it et al.} \cite{moshari2014} numerically investigated the problem of the exit of the horizontal cylinder using a method based on VOF (volume of fluids) from the starting depth until the dewetting of the cylinder. The motion was driven by applying a constant force and the considered starting depths were less than 3 times the cylinder diameter. Both 2D and 3D simulations were performed, which allowed determining the motion of the free interface (2D case) and showing some interesting dewetting phenomena (3D case) along the cylinder after the cylinder crossed the liquid interface. This dewetting could be observed in the case of a hydrophobic coated cylinder for example.

From a physical point of view, the drainage depends on the amount of liquid entrained, and thus on the characteristics of the interface crossing, but also on the flow dynamics during the initial immersed phase (e.g., wake behind the object). The challenge is that the liquid thickness above the object varies by more than six orders of magnitude during the entire process, i.e., the cylinder is first below the surface ($\approx$ m) and the lubrication drainage starts when the thickness of the film is below 100~$\mu$m.
 	
\section{Framework} \label{framework}
The present work aims to provide experimental data on the complete motion of a horizontal cylinder crossing the interface, including (i) the elevation of the fluid surface during the motion of the cylinder inside the bath, (ii) the crossing over, (iii) the entrainment of the fluid, and eventually the link with the drainage of the fluid around the cylinder. The experimental set-up and the choice of the cylinder aspect ratio were designed such as to approach a two-dimensional flow around the cylinder, namely invariant data regarding any translation along the axis of symmetry of the cylinder. This later requirement benefits numerical simulations. 

Two strategies were envisaged to reach 2D conditions of the flow around the cylinder. The aspect ratio $AR$ defined as the ratio between the length $L$ and the radius $a$ of the cylinder, can be increased in such a manner that the flow close to the ends of the cylinder influences less and less the experimental results. The second strategy consists of adding vertical end-plates to the cylinder to force the fluid to flow perpendicularly to its symmetry axis. Based on PIV measurements, we will show that a large $AR$ and the vertical end-plates are required to obtain conditions as close as possible to 2D.  

The starting depth $d$ is an important parameter regarding the development of the cylinder wake. Particle Imaging Velocimetry (PIV) was performed to characterize the fluid motion before the crossing. In the present experimental set-up, the results are independent of the starting depth when $d>12 a$ considering the acceleration distance required to reach a constant velocity. 

The measurements were performed for two fluids: water and silicone oil (50 cSt). The aim of studying such different liquids was twofold: to assess the influence of the fluid kinematic viscosity $\nu$, quantified by the Reynolds number $Re=\frac{2Ua}{\nu}$, but also to address the dewetting around the cylinder. In particular, fast dewetting was observed with water because the cylinder was made out of metal. This could be mitigated with the oil, such that a slow drainage of the entrained fluid around the cylinder could be achieved over a larger range of time. 

In practice, for a given fluid, using the most appropriate cylinder (largest $AR$ with or without end-plates) and starting with a sufficiently large depth, only one parameter is to be tuned: the imposed vertical speed $U$ of the cylinder, reflected by the Froude number, $Fr=U^2/ga$, where $g$ is the acceleration due to gravity.  The time evolution of the height of the fluid interface above the cylinder, of the local film thickness and the force necessary to keep the speed constant were measured for different speed values.

The experimental setup is described in Sect. \ref{experiment}, including the different measured quantities. A typical experiment is also presented to provide a qualitative description of the different regimes. The experimental conditions and limitations can be found in Sect. \ref{limitations}, with a detailed discussion about (i) the size of the cylinder and how to reach the 2D conditions, and (ii) the influence of the starting depth on the wake and the results. It is important to mention that, as already stated, we will show that end-plates and high aspect ratio cylinders are both required to reach 2D conditions. Consequently, most of the experimental results presented in this paper were obtained for the $AR=12$ cylinder equipped with end-plates.  Nonetheless, two experiments were performed under different conditions, namely the film thickness direct measurements ($AR=10$ was used) and the force measurements ($AR=12$ was used but without end-plates). The reasons for these choices are given in the corresponding sections.

The measurements of the interface deformation during crossing are summarized in Sect. \ref{motion}. The upper position of the interface was recorded as a function of the position of the cylinder between an arbitrary starting time (i.e., when the top of the cylinder reaches the position of the interface at rest) and the crossing time (i.e., when the bottom of the cylinder reaches the position of the interface at rest). By image analysis, the maximum height reached by the air-fluid interface was detected as a function of the position of the cylinder during the motion. Knowing the position of both the interface and cylinder, the thickness $h$ of the fluid above the cylinder can be quantified and analyzed. The vertical speed of the cylinder was tuned between 0.1 and 1.0 m/s (Sect. \ref{bulge}).   The final stage of the drainage was studied using a Chromatic Confocal Point Sensor (CCPS, see Sect. \ref{drainage}), which allowed a direct measurement of $h$ as a function of time.  However, the CCPS was used only in the case of silicone oil because the water deweted too quickly on the aluminum cylinders, preventing any meaningful measurements before drying.

During the uniform vertical motion of the cylinder, the force necessary to maintain a constant speed changes because of the time variations of three elementary contributions: drag, buoyancy, and liquid entrainment. The force was measured (Sect. \ref{force}) as a function of the position of the cylinder (or, equivalently, as a function of time) for different speeds $U$ between 0.1 to 1.3 m/s. The cases of water and oil are compared for the largest cylinder aspect ratio considered in this work, $AR=12$, to approach as best as possible 2D conditions.  However, end-plates were not used due to their large influence (friction drag and entrainment of extra liquid) on the force measurements. Finally, global conclusions are drawn in Sect. \ref{conclusion}.

\section{Experimental set-up} 
\label{experiment}

The experimental setup consisted of a large fluid tank fitted with a lifting system that was used to pull or push the cylinder through the air-liquid interface. For the measurements, we used one or two synchronized high-speed cameras, a force sensor, a film thickness measurement device (CCPS), and a PIV system. 

Figure~\ref{fig1} (top) presents three different measurement configurations of the experimental set-up that are developed here below, namely the pushing mode (left), the film measurement (center), and the pulling mode (right). 

The fluid tank was made of glass in the case of water and of PVC in the case of silicone oil with dimensions of $78.5$~cm x $72.5$~cm x $27.5$~cm for the length, height, and width, respectively. The lifting system was composed of a stage that moved along a linear guide. The motion was induced by a toothed belt entrained by a step motor. The cylinder under test for experiments was screwed to the stage via the carbon tube. The axis of symmetry of the cylinder was always oriented parallel to the fluid surface and was fixed on the moving frame by a rod that was screwed at one point of the symmetry plane of the cylinder.    

The frame {\bf pushed} the cylinder upwards so that the fluid surface could only be deformed by the top side of the cylinder. However, in the case of the force measurements, the cylinder was connected to the stage only using one carbon fiber tube and was {\bf pulled} upwards instead of pushed. This allowed us to reduce the influence of the frame (weight and liquid entrainment).   To reach the constant speed set by the user, the acceleration was set to the maximum allowed by the motor, i.e., 4~m/s$^2$. In so doing, the cylinder could cross the interface with a constant speed between 0.1 and 1.3~m/s. The experiments were carried out at a room temperature equal to $20\pm2 ^{\circ} $C. 

The cylinder was made out of aluminum and had a smooth surface.  Two types of cylinders were manufactured. One cylinder family had a radius $a=$25~mm for lengths between 25 and 250~mm corresponding to aspect ratios  $AR$=1, 2, 4, 6, 10. Another cylinder type had a smaller radius $a=$12.5~mm to achieve the largest considered aspect ratio, $AR$=12.  The end-plates were added to the cylinder for some experiments (as specified in the text). The end-plates were 200~mm in diameter and 4~mm in thickness for the cylinders with the larger radius, and 150~mm for $AR=12$ for the same thickness.

The coordinate system ($xyz$) is shown in Fig. \ref{fig1}. The $x$-axis is oriented along the length of the tank, and the $z$-axis is along its width. The origin was set at the ``at rest'' free fluid-air interface considered as the horizontal plane of reference ($xz$). The pushing/pulling $y$-axis passes by the origin and is aligned with the vertical direction. The coordinate $Y_c$ designates the vertical position of the center of mass of the cylinder. The coordinates $y_c=Y_c/a$ are then normalized by the radius of the cylinder, $a$. The position $y_c=0$ is defined as the position of the cylinder for which the cylinder is half plunged in the bath (namely in the $xz$ plane). Consequently, $y_c=-1$ corresponds to the position when the top of the cylinder touches the fluid surface from the bottom; the cylinder is completely above the surface when $y_c>1$.

Two of the controlled parameters in this study are the Froude and  Reynolds numbers, which are varied by changing the cylinder exit velocity and the fluid viscosity. As mentioned earlier, two different kinds of fluids were used, i.e., water and silicone oil (50 cSt, PMX-200). The specific gravity of oil was 0.96 and kinematic viscosity was $5.0\times 10^{-5}$ m$^2$/s at 20 $^\circ$ C.

\subsection{Measurements} \label{measurements}
Different experimental techniques were used for the different stages of the cylinder crossing.
\par {\it Particle velocimetry measurements -} A 2D planar PIV system was used for investigating the flow during the initial phase when the cylinder was still fully immersed. For the PIV exploration, the LASER had a wavelength of 532 nm and a peak power of 4 W and the seeding particles were $\sim$ 20 $\mu$m in size. The post-processing of the PIV images was done with the open-source software PIVLab \cite{thielicke2014pivlab}.

\par {\it Fluid surface deformation -} A white backlight (Effilux LED) was used to illuminate the cylinder and the fluid surface. Two high-speed M-310 phantom cameras acquired images at a rate of up to 3200 Hz. The cameras were perpendicular to the tank so that their respective field of view allowed a complete visualization of the crossing. The air-fluid interface was set in the middle of the image to reduce the parallax and to allow the tracking of the interface position.

\par {\it Force measurements -} A strain gauge SCAIME K25 (20~N) measured the force acting on the cylinder during the whole motion (from the initial acceleration to the end of the drainage). The gauge was calibrated using known weights. The data acquisition was performed using a datalogger (Picolog). The camera trigger signal was also recorded by the Picolog, which allowed the synchronization of the force measurements and the images captured by the camera. The force on the pulling rod was also measured in the absence of any cylinder to subtract the contribution of the rod and the corresponding entrained liquid from the measured force.
\begin{figure}
	\centering
	
	\includegraphics[width=0.9\linewidth]{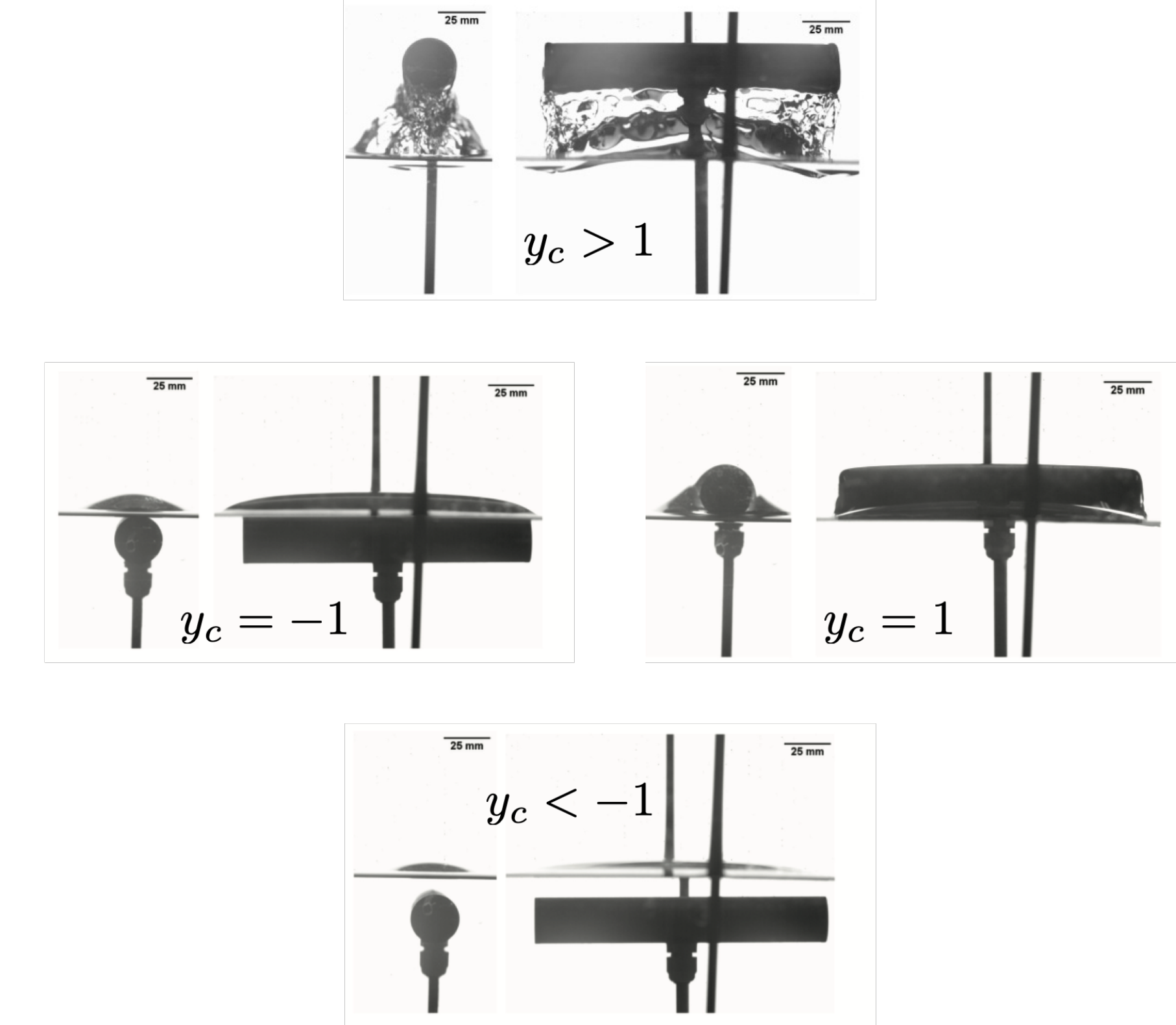}
     \caption{Front and side views of the cylinder moving in the water and exiting the water surface for $AR = 12$ at $Fr = 8.2$.  The origin of time $t=0$ s is when $y_c$=0. The different stages are (i) the motion in the fluid ($y_c<-1$), (ii) the beginning of the crossing when $y_c=-1$ to the end of the crossing when $y_c=1$, and (iii) the liquid falling ($y_c>1$).}
     \label{fig:typical}
	\end{figure}
\par {\it Thinning of the film -} The drainage of the silicone oil located at the top of the cylinder was slow compared to that of the water. Specifically, in this latter case, the film broke up and dewetting was nearly instantaneous after the passage of the cylinder through the interface. On the other hand, the oil wetted well the cylinder. Once the cylinder had exited the bath, a Chromatic Confocal Point Sensor (CCPS, STIL, OPTIMA+) was used to record the thickness of the oil film located at the summit of the cylinder as a function of time. The measurement was achieved when the cylinder was at rest. Consequently, the cylinder had to be stopped in the measurement range of the CCPS. Moreover, the deceleration occurred when the cylinder was far from the interface because we needed to ensure a constant speed during the whole crossing process. For the largest considered speed, the cylinder needed a distance of approximately $17$~cm to accelerate from rest to the aimed speed and the same distance to decelerate to rest. The cylinder was thus stopped about $30$~cm above the surface of the bath to allow the complete crossing of the cylinder before deceleration.

\subsection{Description of a typical experiment}
\label{description}

A typical experiment began by positioning the cylinder at a desired initial position $d$ in the tank. The initial position may vary for each set of experiments according to what was measured and to the presence or not of end-plates. Ideally, $d$ should be as deep as possible to ensure that the target speed was reached as deep below the surface as possible. Once the cylinder settled, a latency time of $15$~min was observed so that all the surface waves were damped.

At this point, the cylinder was moved upward from its initial position at a constant acceleration $\gamma$ of 4~m/s$^2$ until reaching the desired speed $U$. Then, the upward movement continued at a constant speed. For example, it took a distance of about 120~mm for the cylinder to reach a speed of 1.0~m/s. 
When the cylinder moved upward at a constant speed, the system went through different stages, as illustrated in Fig.~\ref{fig:typical}  for a cylinder of aspect ratio $AR$ = 12 and velocity $U$ = 1.0~m/s (i.e., a Froude number $Fr=8.2$) in water.

{\bf Stage 1} starts when the cylinder is far below the surface and initially moves without visibly disturbing the free surface. At a certain depth, the free surface starts deforming: it elevates and presents a visible bump-like profile (see Fig.~\ref{fig:typical}) until the cylinder starts crossing the interface. The thickness $h(y_c(t))$ refers to the vertical distance between the top of the cylinder and the surface elevation at time $t$ (see Fig.~\ref{fig1}(bottom)), and $h^*=h(y_c=-1)$ corresponds to the thickness at the specific moment when the top of the cylinder reaches the surface of the bath at rest. The thickness $h(y_c(t))$ was tracked using high-speed cameras and the flow inside the bath was measured using PIV.

{\bf Stage 2} is the crossing-over, also called the bulging regime. The crossing starts when the top of the cylinder reaches the initial position of the free water-air interface, i.e., at $y_c=-1$ (see Fig.~\ref{fig:typical}). The time $t = 0$~s is set at the moment when the cylinder axis of symmetry crosses the interface ($y_c=0$), such that $y_c=Ut/a$ during the constant velocity phase. The crossing ends when the bottom of the cylinder reaches the initial position of the interface, i.e., at $y_c=1$, and from that moment, the interface profile starts resembling that of the cylinder.

 Finally, {\bf stage 3} corresponds to the waterfall breaking and then ligament fragmentation \cite{wei2024}.
 
 The drainage stage starts immediately after the cylinder top surface crosses the horizontal plane of reference. This means that the drainage starts with stage 2 and is prolonged throughout stage 3. When the cylinder emerges out of the water, the liquid entrained by the front and by the rear of the cylinder forms a cascade (see Fig.~\ref{fig:typical}, $y_c>1$). This cascade continues until all the liquid has drained out.
 

\begin{figure*} 
		\centering
		
	
		
	(a)	\includegraphics[width=0.45\textwidth]{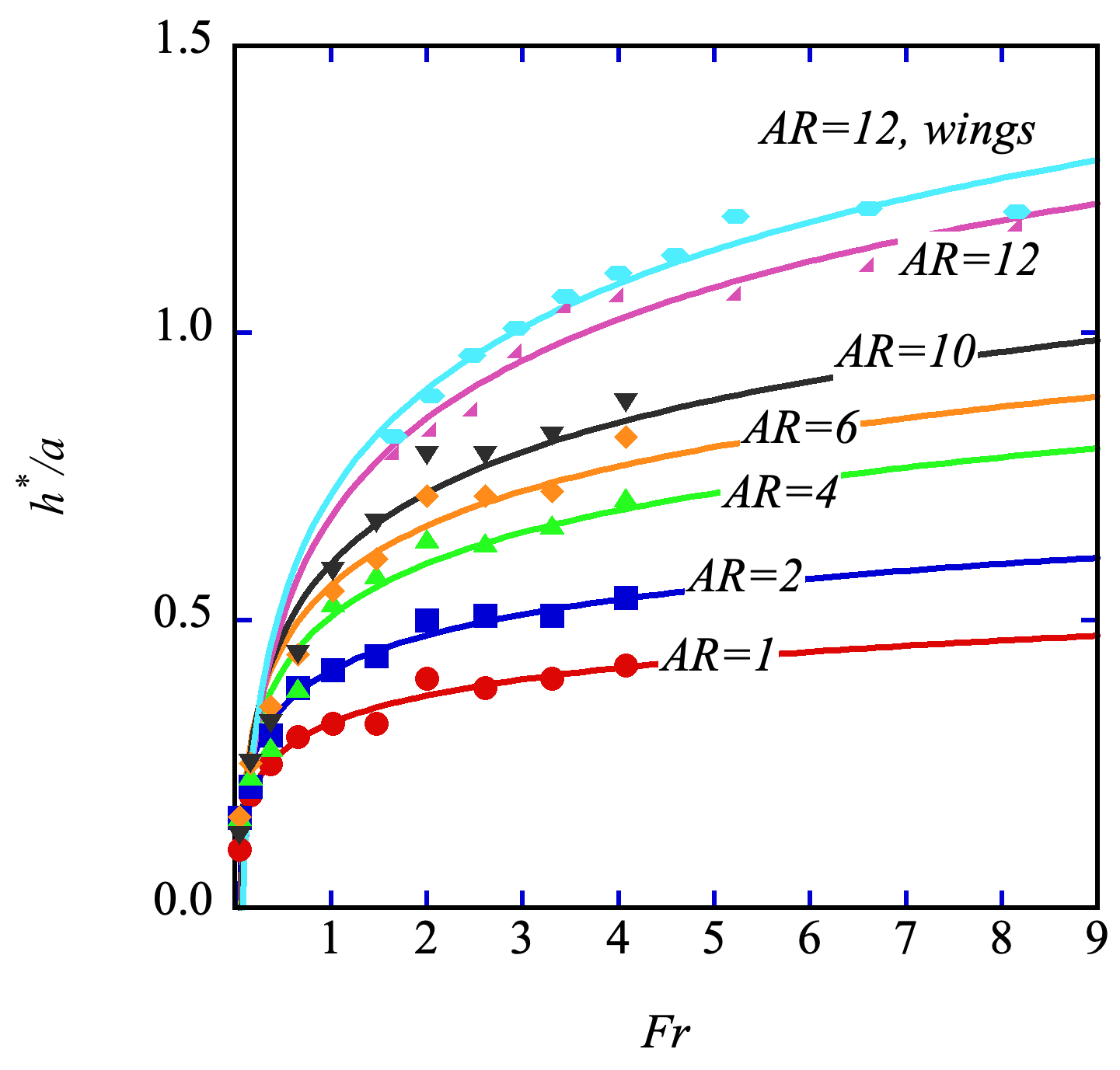}
	(b) \includegraphics[width=0.45\textwidth]{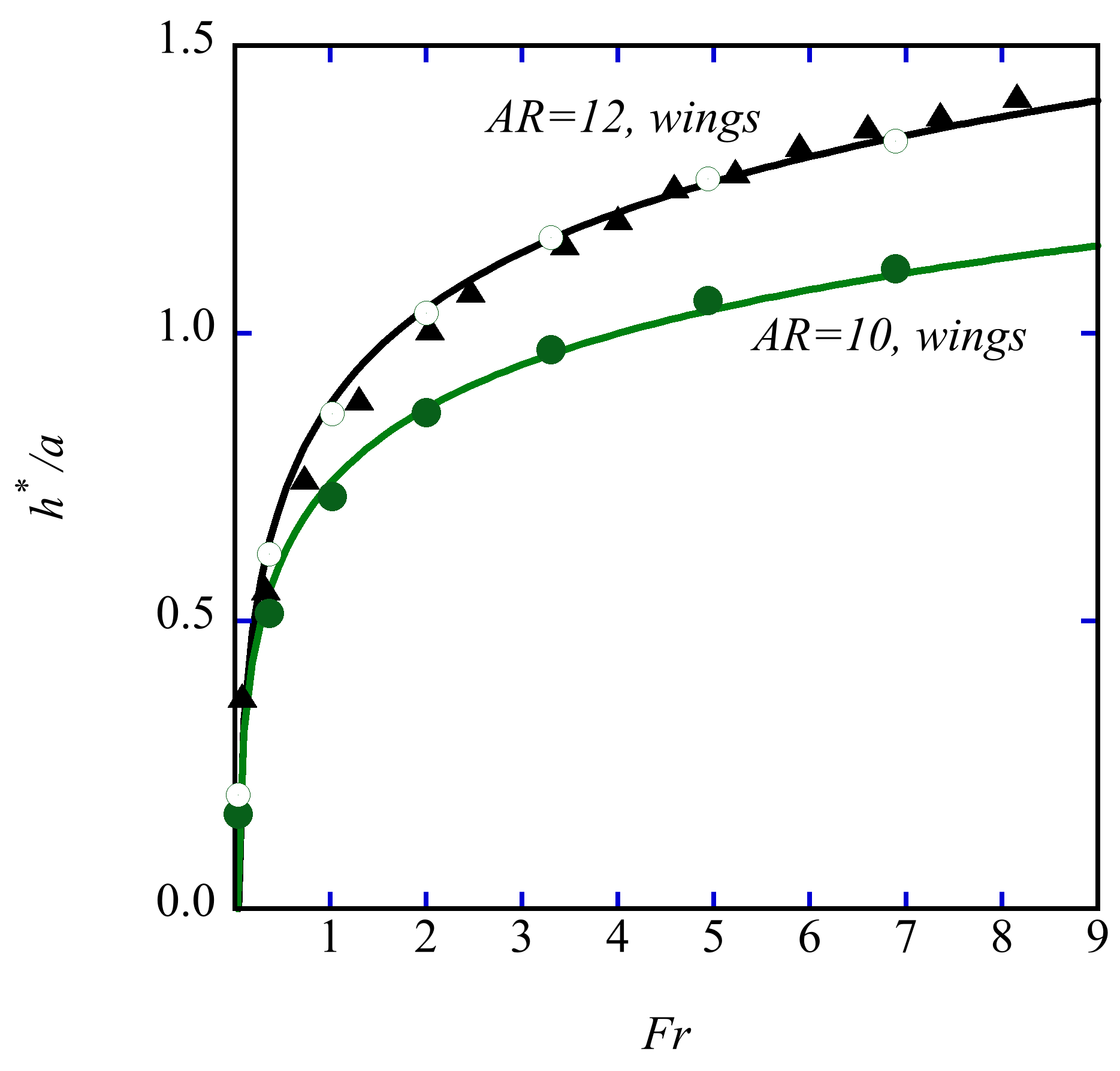}
 
 (c) \includegraphics[width=0.45\textwidth]{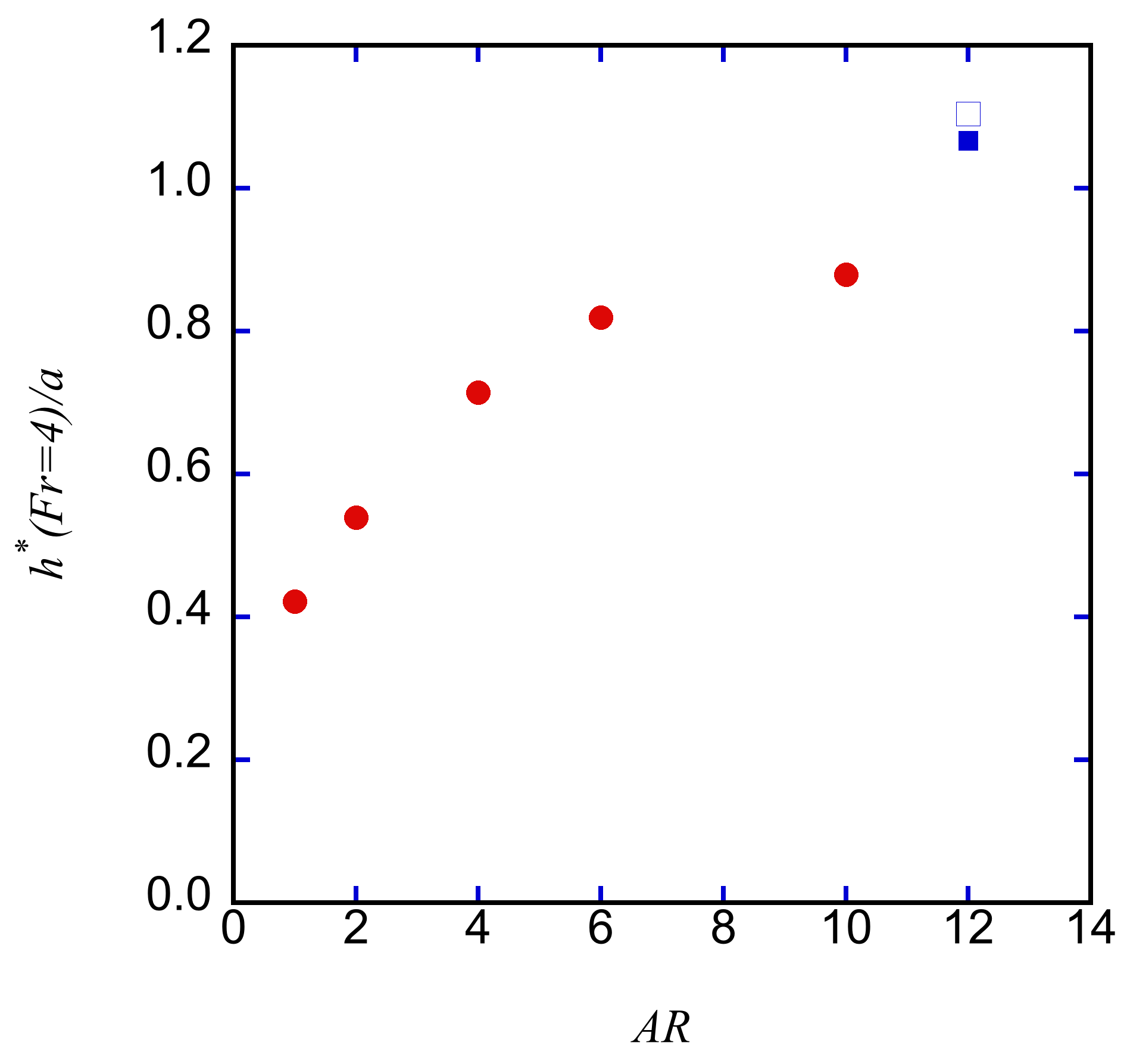}
		\caption{  (a) Water:  normalized thickness $h^*/a=h(y_c=-1)/a$ as a function of the Froude number for an initial depth $d=15a$ and different aspects ratios $AR$. All cylinders are without end-plates except for the largest $AR=12$ (cyan). The different symbols represent the average value of 7 experiments and the lines are logarithm fits (see text). (b) Silicone oil:  same plot as (a) but for two aspect ratios: green bullets $AR=10$ ($d=15a$) and black triangles $AR=12$ ($d=32a$), both cylinders equipped with end-plates. The open circles correspond to the results obtained for $AR=10$ multiplied by 1.2. (c) Water: normalized thickness $h^*/a$ for $Fr$=4 extracted from (a) as a function of the aspect ratio. The red bullets correspond to cylinders with radius $a=25$~mm without end-plates, the solid blue square to the cylinder with $a=$12.5 mm without end-plates, and the hollow blue square to the latter cylinder but equipped with end-plates.}
		\label{fig:ar}
	\end{figure*}

\section{Experimental conditions and limitations}\label{limitations}
We aimed to approach as closely as possible a two-dimensional flow around the cylinder to allow comparison with 2D models. The two strategies followed to achieve this, i.e., cylinders with large aspect ratios and end-plates, are here assessed by analyzing the film thickness as the cylinder emerges out of the bath and the flow lines obtained by PIV measurements. 

Additionally, the impact of the starting depth $d$ is evaluated, accounting for the acceleration (deceleration) length needed to reach constant velocity (rest), but also for the limited size of the tank and the possible presence of end-plates. In particular, the wake behind the cylinder during the immersed phase is characterized through PIV visualization. 

\begin{figure*}
	\centering

	\includegraphics[width=0.43\linewidth]{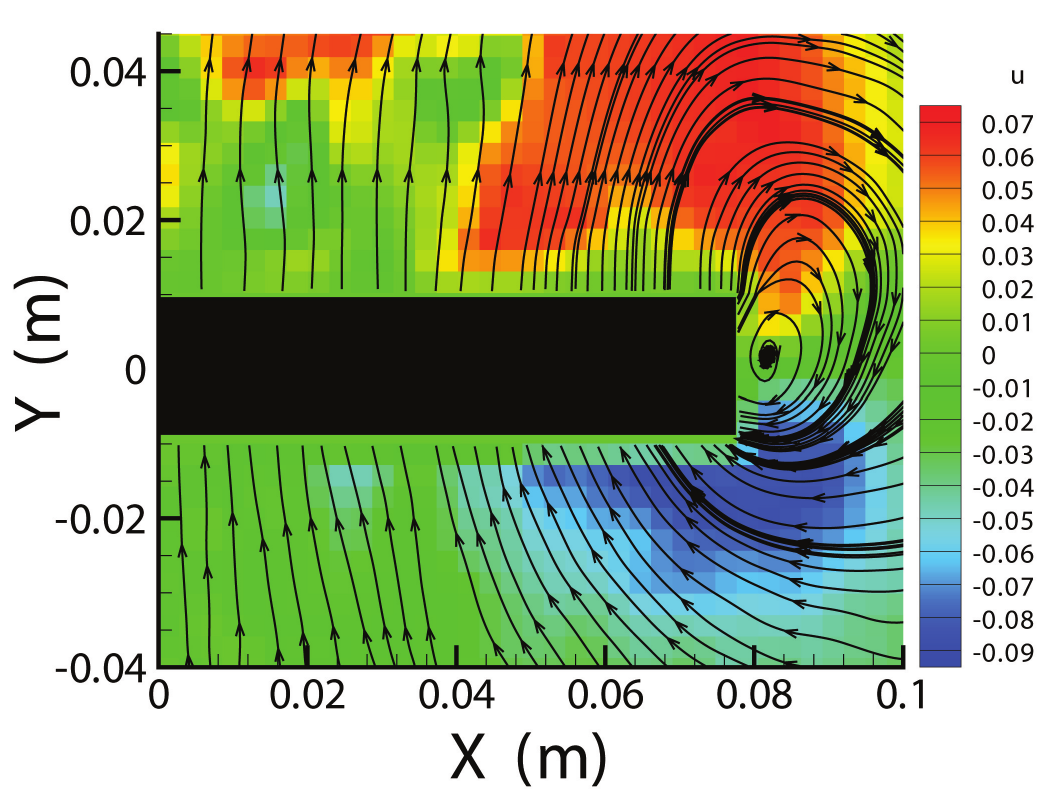}
	\includegraphics[width=0.43\linewidth]{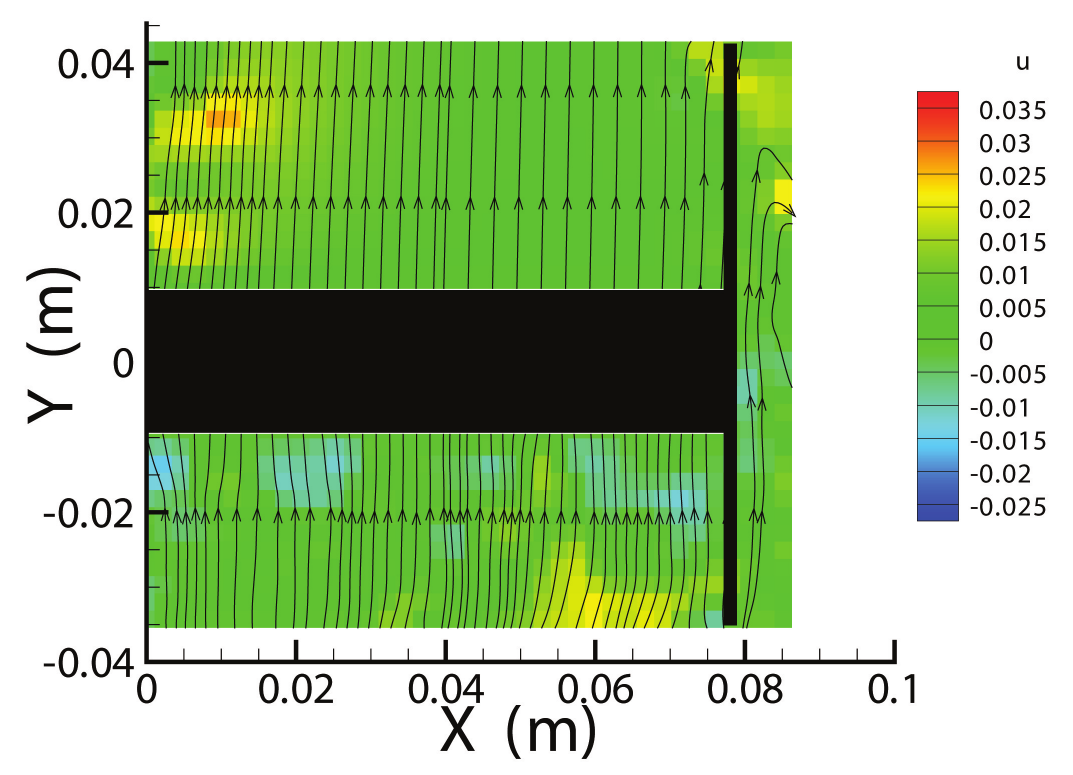}   
	\caption{Water: side view of the flow field ($xy$-plane)  at $U$ = 0.5~m/s and $Fr$ = 2 for the $AR$ = 12 ($a=12.5$~mm) cylinder (a) without and (b) with end-plates. Streamlines in the absolute frame of reference and contour of the $x$-direction velocity. The images show a little bit more than half of the cylinder that is 150~mm long.}	
	\label{fig:piv-endplates}
\end{figure*}
\subsection{2D conditions: aspect ratio}\label{aspect}

An interesting observable is the height $h^*/a = h(y_c=-1)/a$ reached by the fluid interface when the cylinder reaches the interface. For the water case, the measurement of $h^*/a$ is presented in Fig.~\ref{fig:ar}(a) as a function of the Froude number for six cylinders without end-plates whose aspect ratio spreads between 1 and 12. The figure also includes the case of the cylinder with the largest $AR$ equipped with end-plates. The height $h^*/a$, and thus the amount of entrained fluid, increases with the speed of the cylinder, i.e., with the Froude number, and with the aspect ratio. The observed behavior of the entrainment is in line with the observation of Haohao {\em et al.} \cite{haohao2019numerical}, who reported that the waterfall breaking becomes more intense with increasing Froude number. The data were fitted using the logarithmic law $h^* /a=A \ln(Fr)+B$ where $A$ and $B$ are fitting parameters \cite{lionel}. These fitting parameters depend on $AR$ with $A$ between 0.07 and 17 and  $B$ between 0.32 and 0.60. The fits are shown as solid curves in Fig.~\ref{fig:ar}(a). 

The cylinders with $AR\leq 10$ have a radius of 25~mm. For this family of cylinders, we observe that the curves $h^* /a(Fr)$ become closer and closer with increasing $AR$. This is further illustrated in Fig.~\ref{fig:ar}(c), which reports the values of $h^* /a$ for $Fr$=4 as a function of the aspect ratio. The red symbols (corresponding to the cylinders with a radius $a=25$~mm) seem to saturate for large aspect ratios.  One could thus expect that for even larger aspect ratios the curve $h^*/a(Fr)$ would collapse. For comparison, the thickness $h^*/a$ as a function of the Froude number is also shown for the $AR=12$ cylinder, with and without end-plates, but with a radius of 12.5 mm (magenta and cyan data points in Fig.~\ref{fig:ar}(a)). Surprisingly, the two curves significantly depart from that for $AR=10$, even though a saturation regime was nearly reached in this latter case. On the other hand, the thickness $h^*/a(Fr)$ is closer for the two $AR=12$ cylinders with or without end-plate. The values of $h^*/a(Fr=4)$ for $AR=12$, with and without end-plates, are also reported in Fig.~\ref{fig:ar}(c) (hollow and solid blue squares, respectively). The values are about 1.2 times larger than for the cylinder with $AR=10$ and $a=25$~mm. 

These results seem to indicate that, in our experimental setup, 
 the radius of the cylinder does play a role despite the normalization of $h^*$. A possible explanation might be the limited size of the tank, as the ratio between the cylinder radius and the tank lateral length or, more exactly, the ratio between the section surface of the cylinder and the surface of the bath changes. This lateral confinement seems responsible for the absolute value of the liquid elevation. Additionally, the comparison between the $AR=12$ cylinder with and without end-plates illustrates the role of end-plates, even at a large aspect ratio: the fluid cannot escape around the cylinder end surfaces, thereby increasing the amount of liquid above the cylinder, i.e., the thickness $h^*/a$. 

These results suggest that a large aspect ratio and end-plates are necessary to approach two-dimensional conditions. We thus now compare for the oil case the $AR=12$ ($a=12.5$~mm) and the $AR=10$ ($a=25$~mm) cylinders both equipped with end-plates in Fig.~\ref{fig:ar}(b). The thickness $h^*/a$ as a function of the Froude numbers shows the same qualitative behavior as for the water. We again observe that the results depend on the cylinder radius because of the confinement effect. Interestingly, we also find that, by simply multiplying the results obtained for $AR=10$ by the same factor of 1.2 as previously, we obtain the same thickness as for $AR=12$ (open circles in Fig.~\ref{fig:ar}(b)). Overall, we essentially observe that the largest aspect ratio is ideal and that the end-plates do play a role. Moreover, the absolute value of $h^* /a$ changes by a multiplicative factor when the radius of the cylinder is changed. This correction for the finite size of the tank is however difficult to predict {\it a priori}. In our case, we found a factor of 1.2 on $h^*/a$ when doubling the radius.

\subsection{2D conditions: end-plates}\label{endplates}

PIV measurements were performed during the cylinder motion below the free surface to determine whether the flow around the cylinder equipped with end-plates is indeed two-dimensional. The most favorable aspect ratio was investigated namely $AR=12$. The velocity field obtained by PIV is presented in Fig.~\ref{fig:piv-endplates}(a) for the cylinder without end-plates and in Fig.~\ref{fig:piv-endplates}(b) with end-plates. Without end-plates, while the flow in the middle of the cylinder is mostly parallel and straight, it starts to exhibit a non-zero horizontal component when approaching the cylinder extremities. The flow is thus not fully two-dimensional, except close to the center of the cylinder. With end-plates, the streamlines are parallel along the whole length of the cylinder, as shown in Fig.~\ref{fig:piv-endplates}(b). The water located above the cylinder cannot escape by the lateral side and two-dimensional conditions are well approximated. 

Consequently, all results reported in the following are based on experiments with cylinders equipped with end-plates, except in section~\ref{force} for the force measurements to avoid a significant bias of the measurements due to the weight of the end-plates and the liquid entrainment by friction on their large surface.  

\subsection{Starting depth}
\label{depth}
\begin{figure*}
	\centering
	
(a) \includegraphics[scale=0.4]{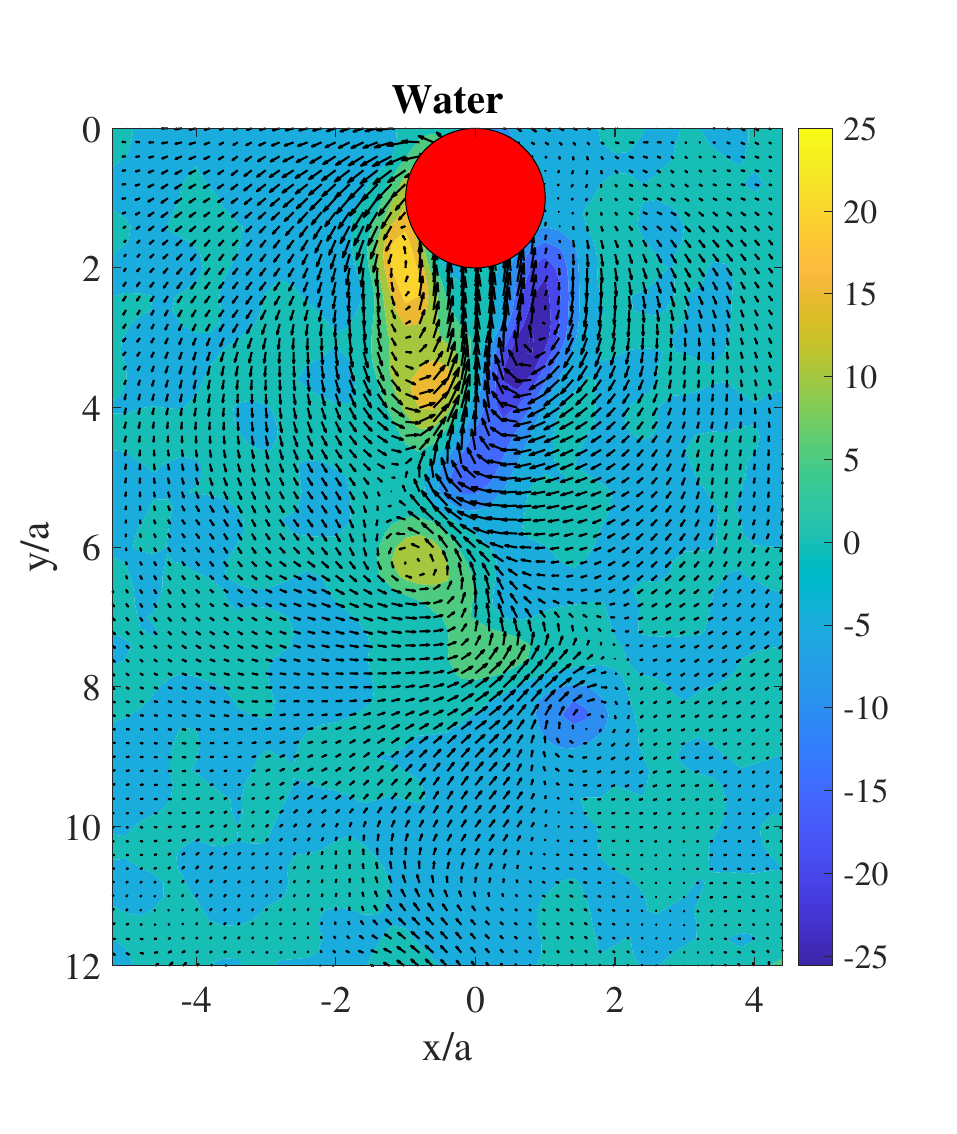}%
(b)	\includegraphics[scale=0.4]{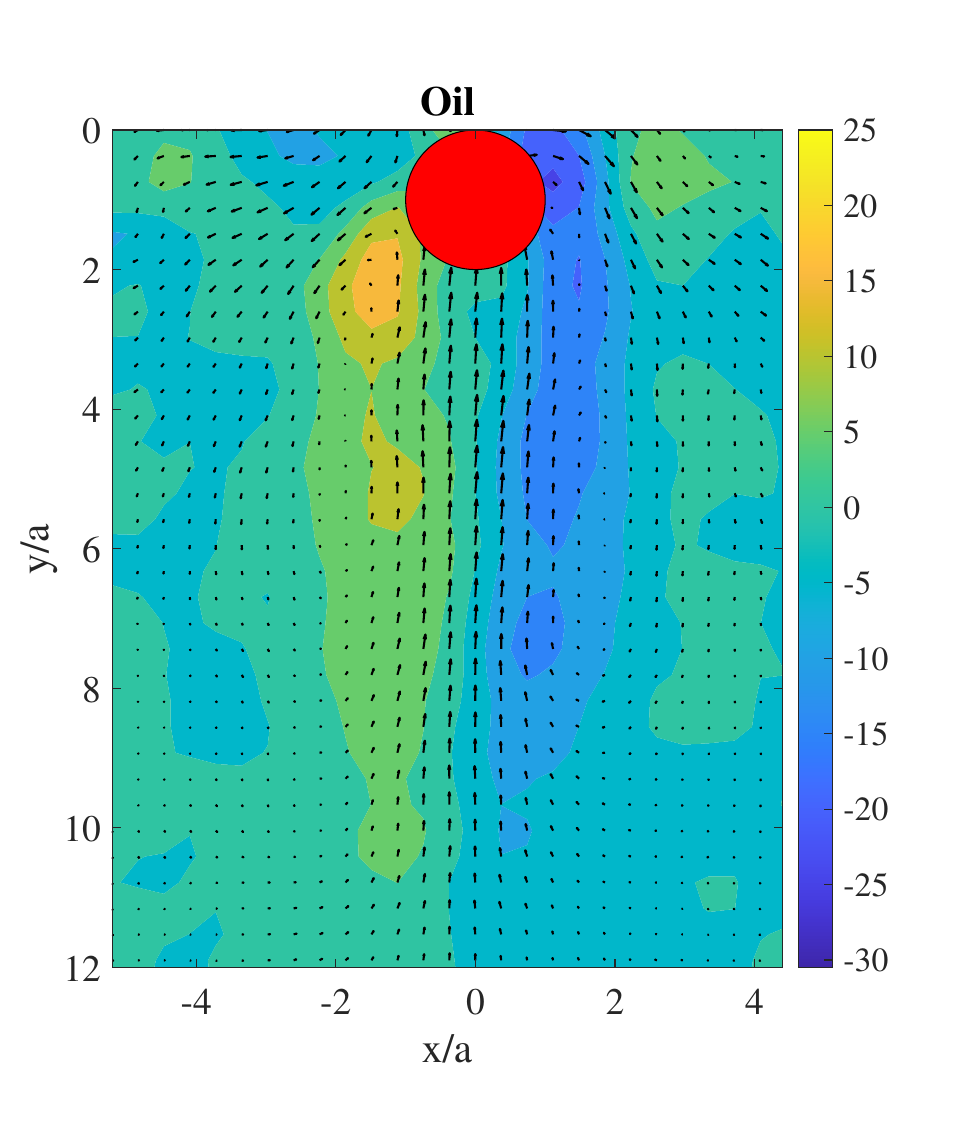}
	\\
(c) \includegraphics[scale=0.4]{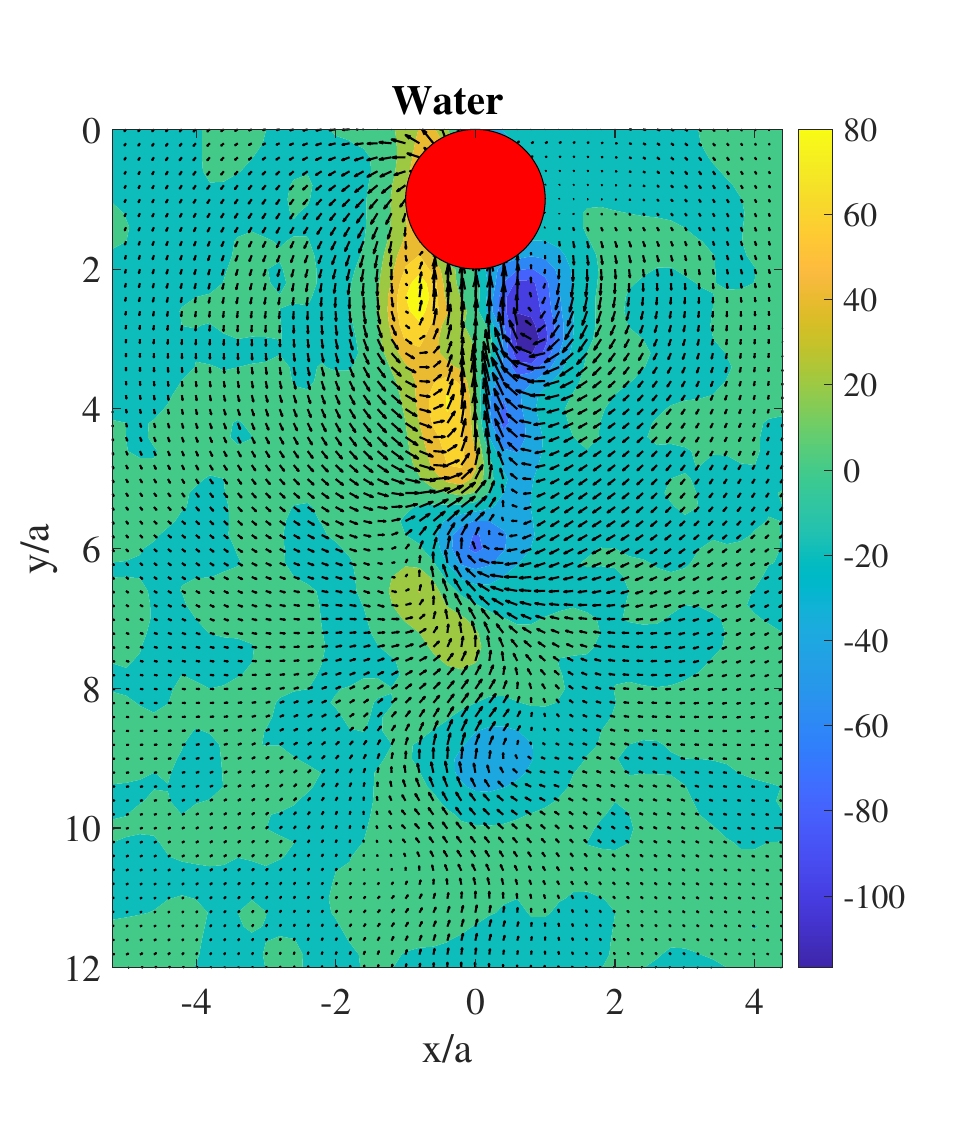}%
(d) \includegraphics[scale=0.4]{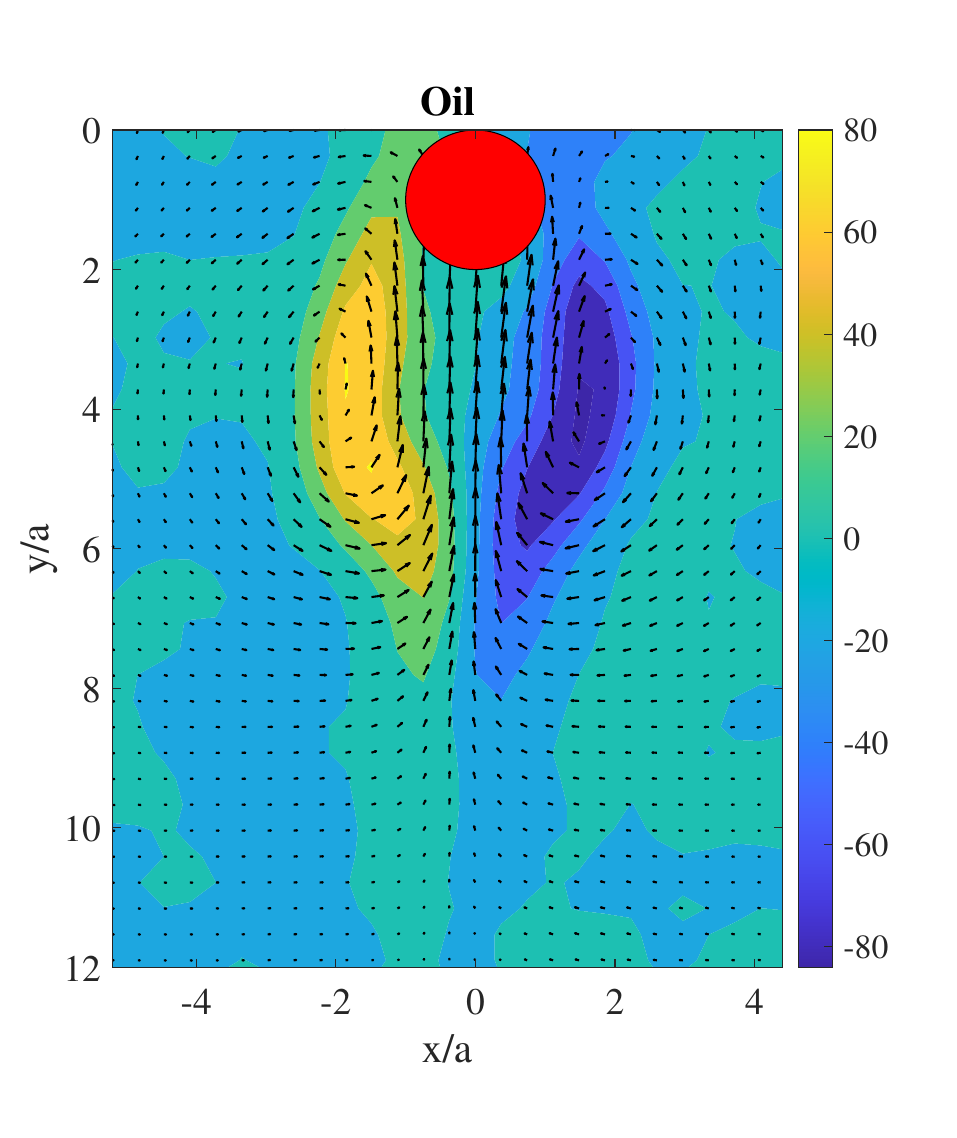}
	\qquad
	\caption{Velocity vectors and vorticity contour in the $xy$-plane through the cylinder mid-point from PIV measurements for the $AR=12$ cylinder with end-plates and a release depth $d=24a$ in water (left: a,c) and oil (right: b,d) at $Fr=0.33$ (top: a,b) and $Fr=8.5$ (bottom: c,d).}
	\label{fig:piv}
\end{figure*}

For the comparison between the motion in the water and in oil, the cylinder with aspect ratio $AR$=12 ($a=$12.5 mm) and end-plates was released from a depth $d=24a$. Figure~\ref{fig:piv} shows the velocity vectors and vorticity contours from PIV measurements for two typical speeds corresponding to Froude numbers $Fr$ = 0.33 (top) and 8.15 (bottom). The Reynolds numbers for the water case (left) are  $Re=5000$ and $Re=25000$, while for the oil case (right) $Re=100$ and $Re= 500$, respectively.  In the case of water, vortex shedding occurs even at a small speed ($Fr=0.33$), breaking the symmetry of the flow. In particular, a pair of clockwise and anti-clockwise vortices are located close to the cylinder and some weaker vortices previously shed in the wake are visible. As expected, the strength of these vortices increases with the cylinder velocity, and thus with the Froude and Reynolds numbers. The symmetry breaking influences both the entrainment and the collapse of the liquid behind the emerging cylinder. The situation is quite different in the case of the much more viscous silicone oil: there is no vortex shedding with a release depth of $24a$ and the wake remains symmetrical even for the most unfavorable sets of parameters considered here (highest speed and deepest possible starting depth, see Fig.~\ref{fig:piv}(d)). The volume of liquid that is perturbed by the passage of the cylinder is thus rather well-defined compared to the case of water. Note that the symmetry of the wake is important for any theoretical development. Furthermore, it has been shown that the evolution of the entrained liquid ahead of the cylinder is related to the presence of the wake \cite{lionel}. 

\begin{figure*}
	\centering
	
		
	(a) \includegraphics[width=0.45\linewidth]{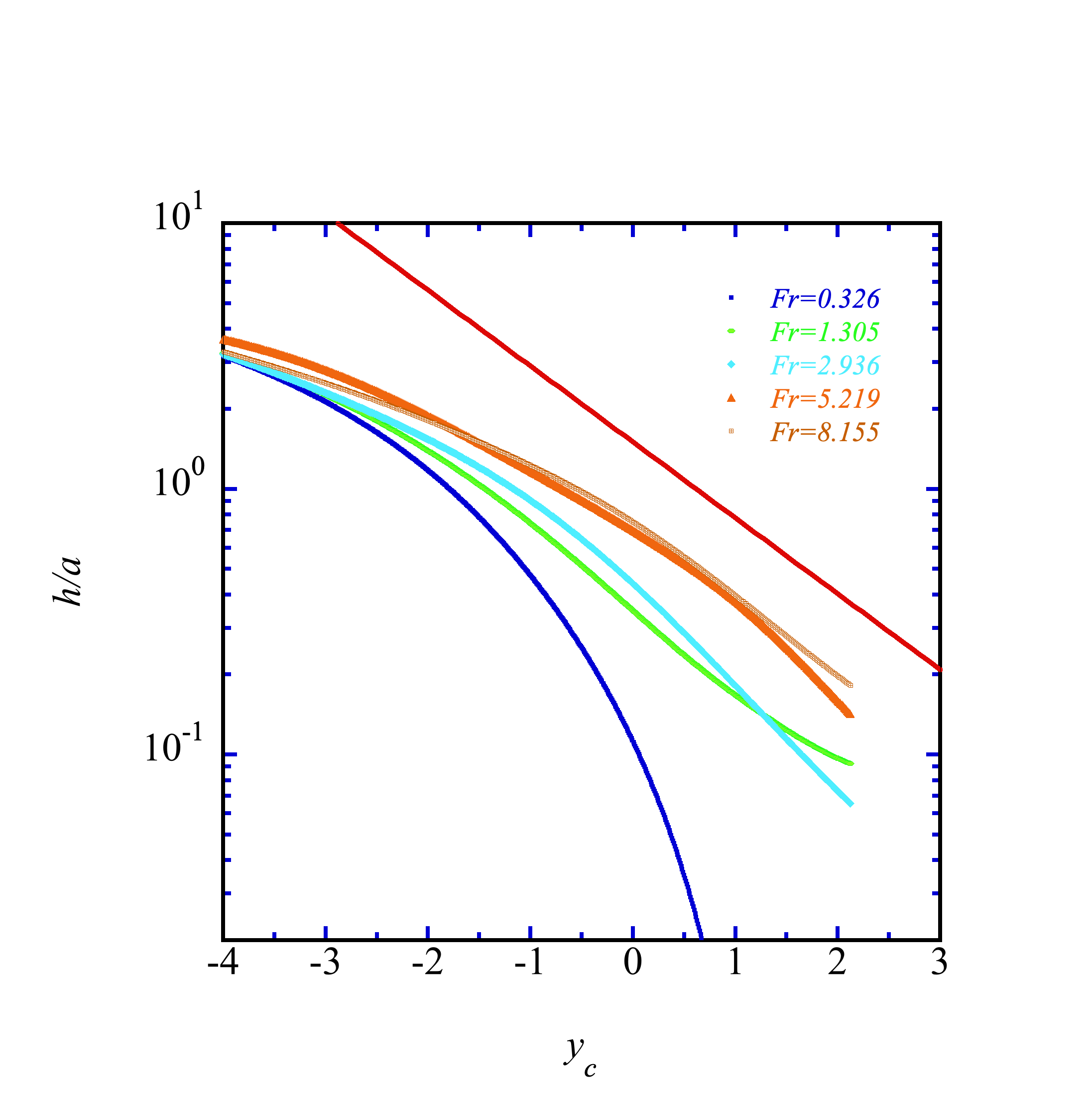}
	(b) \includegraphics[width=0.45\linewidth]{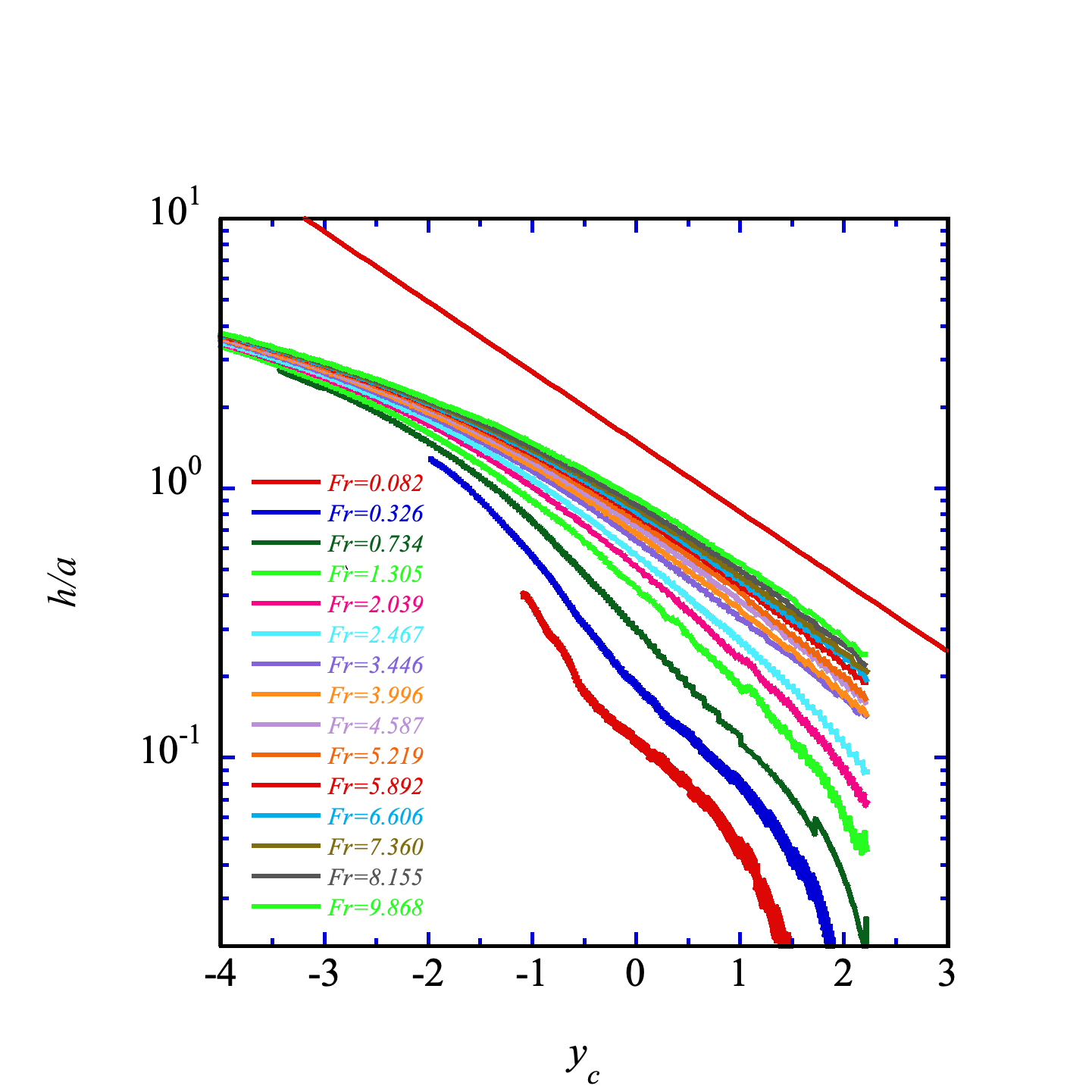}

	\caption{Normalized thickness $h/a$ of the liquid above the cylinder as a function of the cylinder position $y_c$ for different Froude numbers; cylinder with $AR=12$ and end-plates in (a) water for $d = 24a$ and (b) silicone oil for $d=32a$. The red lines represent an exponential decay function with a characteristic time $\tau=19$~ms; to convert the time into $y_c$ units, the exponential functions were calculated for the higher $Fr$ in both cases, which corresponds to $U=1.0$~m/s and 1.1~m/s for water and oil, respectively. }
	\label{fig:h(y)}
\end{figure*}

Overall, the starting depth $d$ does not influence the results concerning the deformation of the interface as soon as the depth is larger than a given value (3 times the diameter of the cylinder according to Liju {\it et al} \cite{liju2001}).  From a practical point of view, we considered the release depth of at least $d>12 a$. This condition is stronger than the theoretical condition but we have to account here for the acceleration phase of the cylinder. Therefore, the cylinder's initial depth was set as deep as possible for each presented case. At last, it is important to mention that the maximum release depth was constrained by the limited size of the tank and by the presence of the end-plates. For a sufficiently large $d$, one could thus expect a destabilization of the wake and a subsequent vortex shedding to also occur for the oil case at the lower Reynolds number.

\section{Motion of the interface}\label{motion}
The evolution of the thickness $h$ between the top of the cylinder and the air-fluid interface was measured as a function of the normalized position $y_c$ of the cylinder for different speeds (pushing mode) between 0.1 and 1.0~m/s. The motion of the interface was obtained by determining the higher position reached by the liquid while the cylinder was moving. The optical axis of the camera was horizontal and contained in the interface plane. In so doing, the interface appeared in the middle of the image. Moreover, the camera was tilted by 90$^\circ$ to have a maximum of pixels along the vertical direction since the sensor had a 10:9 ratio. 

Before the experiments, the cylinder was placed just under the interface, and its position was recorded to determine the coordinate $y_c=-1$ on the image. Then, two pictures of the cylinder were taken above and under the interface to obtain the calibration pixels to millimeters (which differs due to the water or oil refraction). The images were recorded close to the center of the cylinder to reach 2D conditions, even in the presence of end-plates.

The contrast between the interface and the cylinder was sufficient to obtain the position of the interface and the cylinder by subtracting an image of the background, i.e., in the absence of the cylinder. The image resulting from the subtraction was then thresholded. Note that the position of the cylinder was obtained by tracking the lower position of the cylinder (the apex could not be tracked as easily because of the interface deformation when the cylinder approached the surface). As a consequence, the position of the cylinder could only be measured when the bottom part of the cylinder was located in the bath. The position $y_c$ of the cylinder was then extrapolated based on the position of the cylinder before $y_c=0$. The linearity of $y_c$ with time was also checked and the slope allowed to confirm that the set-point speed was correct within a few percent. 

In the case of silicone oil, measurements of $h(t)$ could be acquired over a long period of time thanks to the Chromatic Confocal Point Sensor (CCPS). In particular, because the oil wetted the cylinder, the entrained liquid slowly drained until reaching a thickness lower than 10~$\mu$m. Synchronizing the data obtained by the CCPS with $h(t)$ obtained by image analysis during the interface crossing, we could obtain the history of the film thinning over several time decades. The synchronization was achieved by triggering the data logger with the same signal as the high-speed camera. The recorded images allowed us to determine the time at which the cylinder crossed the interface.

\subsection{Bulging stage ($y_c<-1$)}\label{bulge}

Figure~\ref{fig:h(y)} shows the thickness $h(y_c)$ for the water and the oil ($AR=12$ equipped with end-plates) for several speeds between 0.1~m/s and 1.0 m/s as indicated by the Froude number. The starting depth was $d=24a$ in water and $32a$ in silicone oil. In both cases, the same behavior is observed. The fluid thickness above the cylinder decreases monotonically as the cylinder approaches the air-fluid interface at $y_c=0$. The decrease becomes more rapid when the cylinder approaches the interface and crosses it. Because of liquid entrainment, the interface crossing becomes smoother with increasing Froude number. The curves tend also to be closer and closer to each other. This convergence towards a limiting curve is to be correlated with the dependence of $h^*/a$ on the Froude number $Fr$ (see Fig.~\ref{fig:ar}). These results suggest that the decrease of the thickness follows an exponential decay as a function of the position (i.e., as a function of time since $y_c(t)=U t/a$). 

The exponential behavior can be rationalized through a simple model.  When the cylinder starts crossing the interface, the liquid layer surrounding the cylinder has a thickness $h$ which depends on the polar coordinate $\theta$ (angle formed by the apex of the cylinder, the center of the cylinder, and the position along the cylinder surface).  Because of symmetry, the azimuthal velocity $v_\theta$ of the fluid vanishes at the cylinder apex ($\theta=0$) and increases with $\theta$; the maximum speed $v_m$ is thus expected at $\theta=\pi/2$ (at least for $y_c\geq 0$). The volume $V$ of liquid around the top half of the cylinder is given by $V\approx \pi a L h (\theta)$ and the flow rate $\dot{V}=2 L h(\theta) v_\theta(\theta=\pi/2)$.  If the viscosity is neglected, the liquid is accelerated downwards by gravity and only the geometry of the flow is to be taken into account. We then find that $\dot h \propto h$, at first approximation, namely the thickness $h$ experiences an exponential decay with time. The decay of $h$ as a function of time was fitted by a decreasing exponential $\exp(-t/\tau)$. The red lines drawn in Fig.~\ref{fig:h(y)} correspond to the function $A_0 exp(-y_c a/(U\tau))$ where $A_0$ is an arbitrary constant and $\tau$ is the characteristic time. The characteristic time was found to be 19~ms (in both water and oil cases) and the reference speeds used to draw the lines were chosen as the highest speeds represented in Fig.~\ref{fig:h(y)}, namely $U=1.0$~m/s and $U=1.1$~m/s for water and oil, respectively. Finally, the characteristic time can be related to the ``escape'' speed of the fluid from the top of the cylinder. To a first approximation, we may compute the time $\tau_\mathrm{fall}$ needed to fall from a height equal to the cylinder diameter: $\tau_\mathrm{fall} \approx \sqrt{2a/g} \simeq$ 50~ms. Note that this approximation overestimates the typical characteristic time since the liquid can escape by both sides of the cylinder. 

%
%
%

\begin{figure}
	\centering
	
	\includegraphics[width=0.45\textwidth]{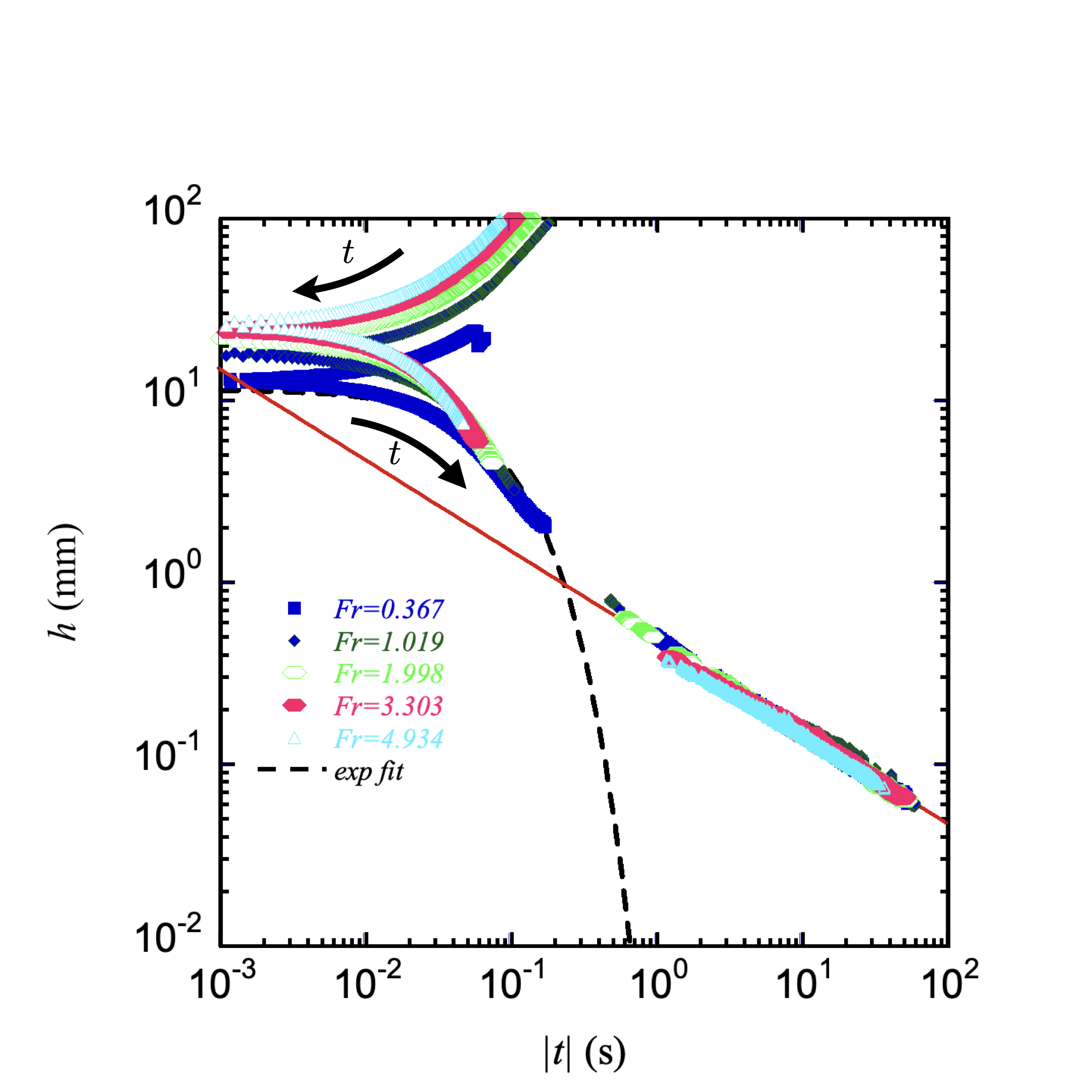}
	\caption{Silicone oil: thickness $h$ as a function of the absolute value of the time for different Froude numbers ($AR=10$ with end-plates, $d=15a$). The data above $h=1$~mm was obtained using image analysis while the data below $h=1$~mm was obtained using the CCPS. The arrows suggest the time direction. The continuous red line is a fit $t^{-1/2}$ of the data obtained with the CCPS; the dashed black line is the fit of the data corresponding to $Fr=0.367$ for $t>0$ by an exponential decay (see text).
	}
	\label{fig:whole}
\end{figure}

\subsection{Drainage stage ($y_c>-1$)}\label{drainage}
When the cylinder emerges out of the liquid, the entrained liquid drains. This process depends on the wetting properties of the liquid on the cylinder material. The water was found to dewet very quickly and the water film broke as soon as the cylinder was above the surface. On the other hand, in the case of oil, the liquid flew down from the top to the bottom of the cylinder forming a cascade back to the tank. The drainage time was much longer than with the water because (i) the oil wets the aluminum, and (ii) the viscosity of the oil is 50 times higher than that of the water. Therefore, it was possible to directly measure the oil film thickness using a Chromic Confocal Point Sensor. For this purpose, the cylinder was stopped at a precise position such that the top of the cylinder was located in the measurement range of the CCPS (this range is as large as $1.3$~mm). Then, the CCPS measured the film thickness as a function of time (acquisition frequency of 2500~Hz). Knowing the deceleration rate, the maximum speed reached by the cylinder, and the time at which the cylinder crosses the interface, we synchronized the data obtained by image analysis and the direct thickness measurements.

Unlike the previous sections, data are here given for a cylinder with aspect ratio $AR=10$ ($a=25$~mm), instead of $AR=12$ ($a=12.5$~mm), equipped with end-plates. The reason is that measurements using the CCPS are much more precise when the radius of curvature of the cylinder is larger. Although the pulling system allows precise vertical positioning of the cylinder, vibrations, and oscillations of the system induce small lateral displacements that cause errors in the film thickness measurement. These errors can be reduced with a larger radius of curvature.

\begin{figure*}
	\centering
	(a) \includegraphics[width=0.45\linewidth]{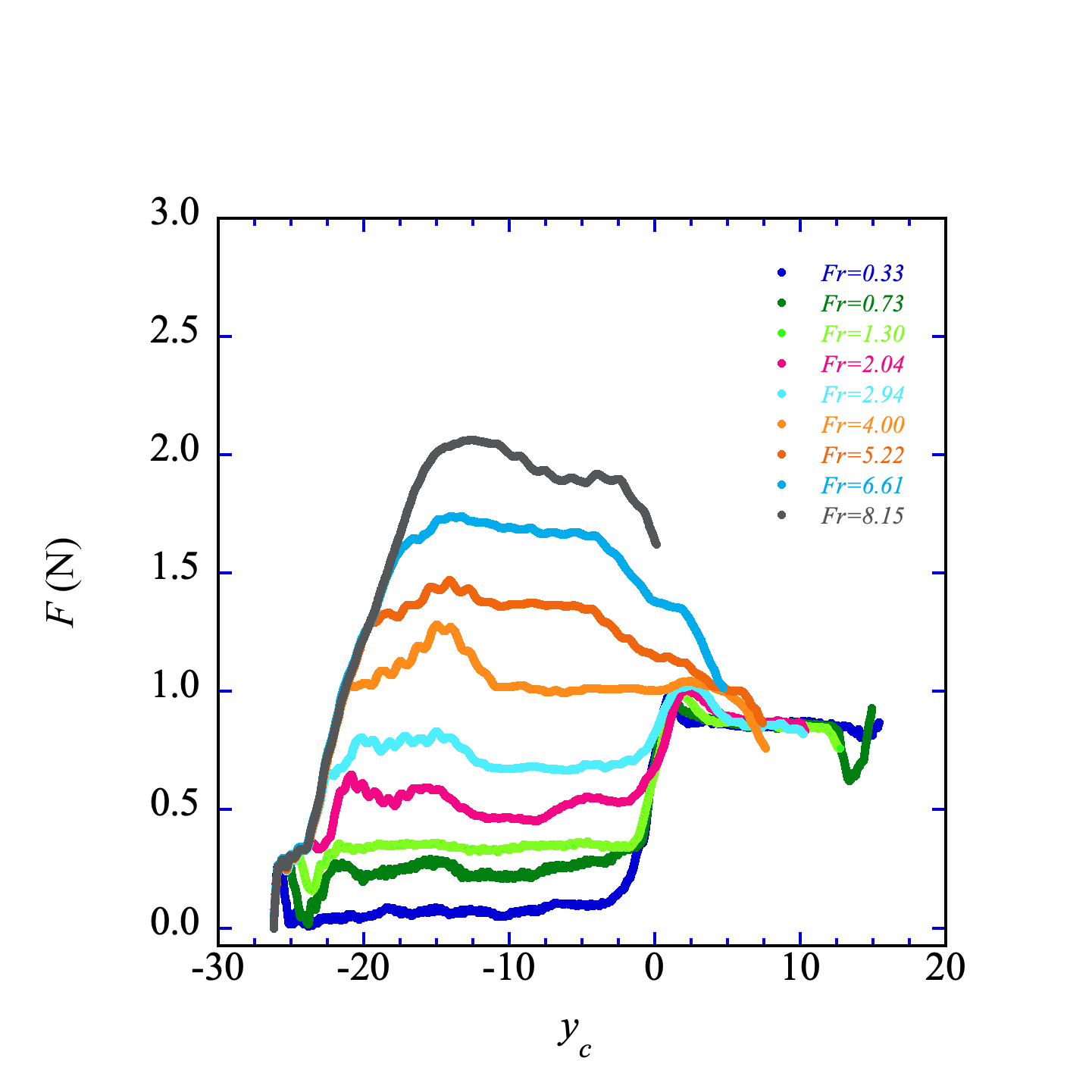}
(b) \includegraphics[width=0.45\linewidth]{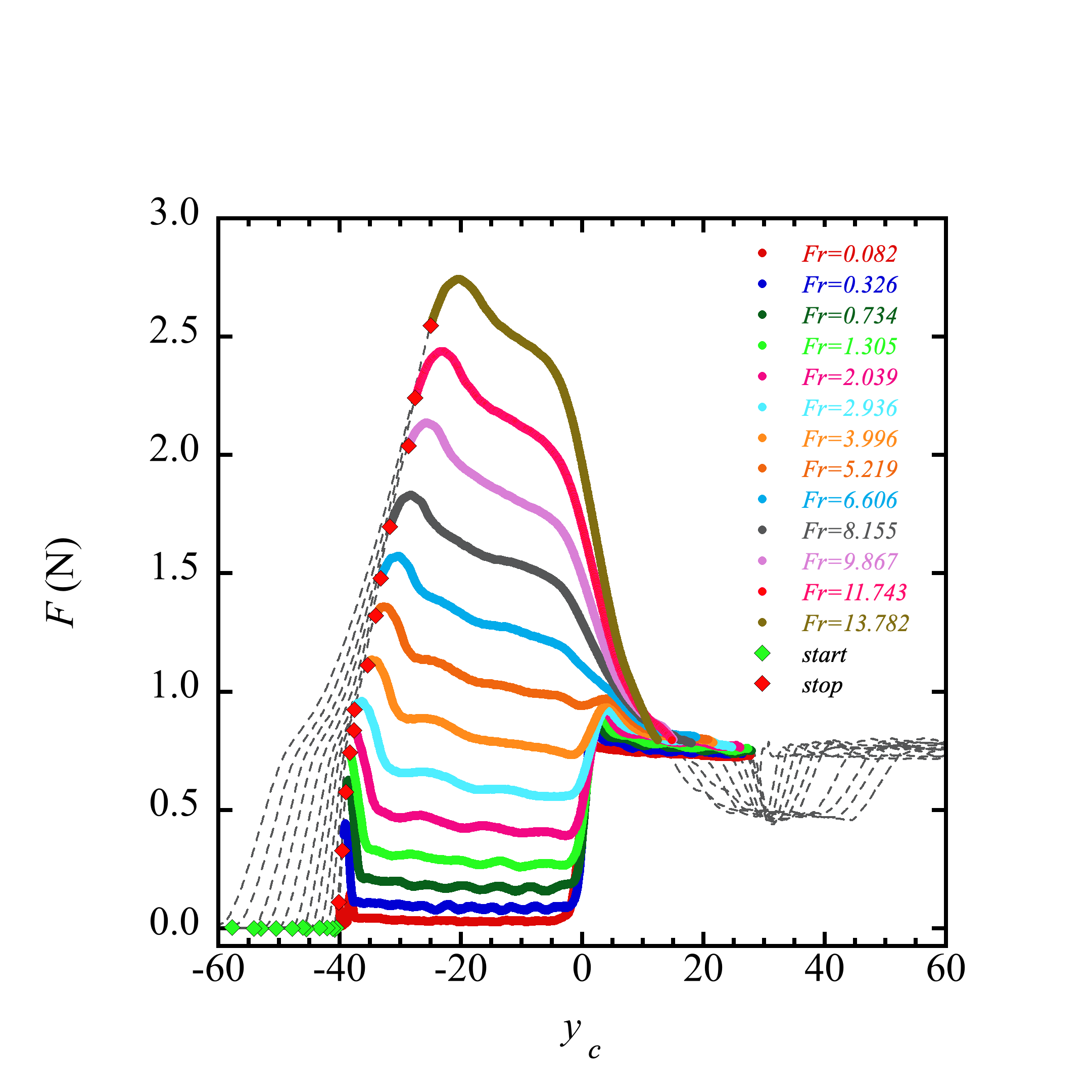}

	\caption{Force applied to the $AR=12$ cylinder without end-plates during its motion as a function of its position $y_c$ for different Froude numbers (a) in water from $d=24 a$, and (b) in silicone oil from $d=38a$. In (b) the acceleration and deceleration phases are shown as dashed grey lines and do not correspond to the actual $y_c$ since during the acceleration $y_c\neq U t/a$. Additionally, the beginning and the end of the acceleration are indicated by the green and red diamonds, respectively. }
	\label{fig:force}
\end{figure*}

In Fig. \ref{fig:whole}, the thickness $h$ from both sets of data (image analysis and CCPS) is presented as a function of time for different Froude numbers. The range of thickness measurements and the range of times considered is large. Because $t<0$~s for $y_c<0$, the absolute value of time is presented in logarithmic scale so that $h(|t|)$ is represented by cusp curves; the upper part corresponds to $h(t)$ when $t<0$~s and the part below the cusp to $t>0$~s. The data for times larger than 1~s were obtained through the CCPS. The superposition of the data from the image analysis and the CCPS provides the variation of the liquid thickness over five decades in time and nearly four decades in thickness. The data from the CCPS are found to be the natural prolongation of the data obtained by image analysis. Moreover, the drainage curves overlap for $t>1$~s, indicating that the drainage dynamics does not depend much on the initial conditions, i.e., the cylinder velocity $U$ here. This confirms the fact that the initial thickness of the draining film above an object is not relevant. The red line in Fig.~\ref{fig:whole} is a fit of the CCPS data using a power law of the time, $h\propto t^{-1/2}$, while the black dashed line is an exponential decay fit of $h$ at the first instant after the interface crossing.
%
A change of regime between the exponential decay and the power law occurs. The exponential decay (when $t>0$~s) is characterized by a characteristic time $\tau=$ 93~ms (in the case of the cylinder $AR=10$, $\tau_\mathrm{fall}\simeq 71$~ms). The regime change occurs at approximately $t\simeq 250$~ms and $h\simeq 1$~mm, and reflects the fact that the fluid viscosity becomes dominant when the film thins. Remarkably, irrespective of the crossing speed, the change of behavior occurs almost at the same time and for the same critical thickness.

\section {Force measurements} \label{force}

The force measurements allow a global measurement from the start to the complete interface crossing. When the cylinder is below the surface, the force results from the contributions of the cylinder weight, the buoyancy, and the drag. When the cylinder is out of the liquid, the buoyancy and drag vanish but the weight of the entrained liquid adds to the cylinder weight. Moreover, inertial contributions are also present during the acceleration and deceleration phases.

\begin{figure}
	\centering

	\includegraphics[width=\linewidth]{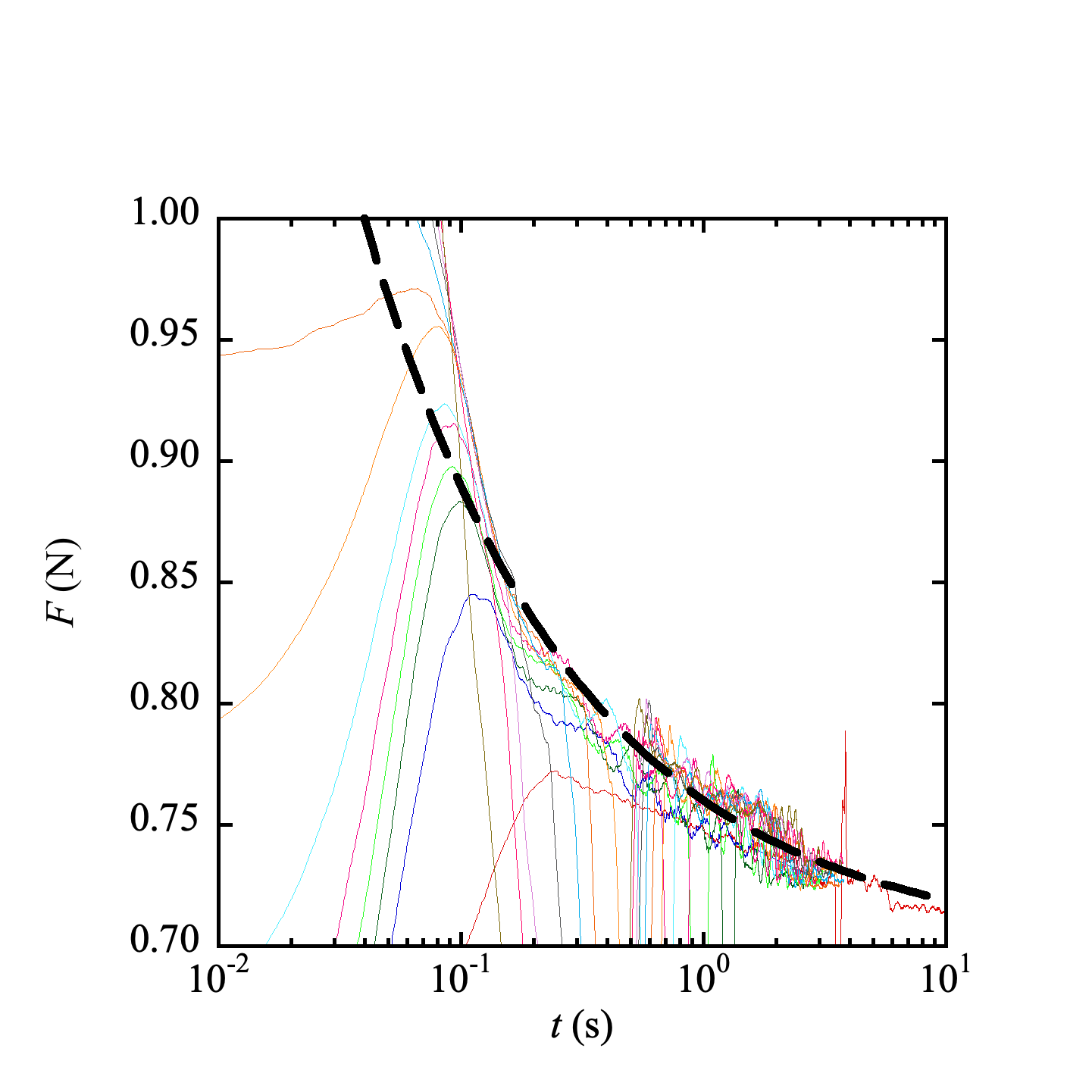}


	\caption{Force applied to the cylinder as a function of time (same conditions as Fig.~\ref{fig:force}(b) for $t>0$~s using a semi-logarithmic scale). The dashed curve represents a $t^{-1/2}$ trend.  
    }\label{fig:zoom}

\end{figure}
The evolution of the force in water and in oil is reported as a function of the cylinder position for different Froude numbers (pulling speeds) in Fig.~\ref{fig:force}. The aspect ratio used for these measurements was restricted to the highest considered value of 12 since the 2D conditions are more closely satisfied when the cylinder is long. Note that the end-plates cannot be used during the force measurement experiments because they can entrain large quantities of oil (Landau-Levich pulling film along a vertical plane) and, in the case of water, generate waves at the surface of the water when crossing the interface (thus before the cylinder interface crossing).

Experimentally, the signal from the data logger provides the curve $F(t)$ (see Fig.~\ref{fig:zoom}). The reference force $F=0$~N corresponds to the measured force when the cylinder is at rest and under the liquid surface. To compare the force measurements for the different speeds considered, the time series are converted into the position of the cylinder: a simple transformation following the rectilinear motion was applied, namely $y_c=Ut/a$, with $t=0$~s when the cylinder is located at $y_c=0$. This means that, during the acceleration and deceleration phases, the conversion $t\rightarrow y_c$ following the linear law does not hold. We choose nonetheless to present the curves including this bias to show the force variation during the complete motion of the cylinder. In Fig.~\ref{fig:force}, the data points corresponding to the acceleration and deceleration phases have been suppressed in the case of the water. In the case of oil, the acceleration phase is delimited by a green diamond when the acceleration stage starts and a red diamond when it ends. Both acceleration and deceleration phases are indicated by grey dashed lines. 

Figure~\ref{fig:force} clearly shows that, until the later drainage phase, the force level strongly increases with the cylinder speed (Froude number), as expected. We can also identify different stages in the evolution of the force. The force first strongly increases during the acceleration phase (dashed line at low values of $y_c$ for the oil) and reaches a maximum shortly after the end of the acceleration. Then, the force starts dropping before reaching a plateau (or a region of lower decreasing rate at high $Fr$ for the oil case). This plateau is particularly marked at low Froude numbers ($Fr \lesssim 4$) and extends until shortly before the cylinder reaches the interface ($y_c < -1$). This plateau region is representative of the drag force acting on the cylinder in the liquid bath. In particular, the magnitude $F_d$ of the force was evaluated when the plateau was visible and was found to be proportional to the Froude number. This corresponds to a dependence on the square of the motion velocity, $U^2$, consistent with hydrodynamic drag. 

 Approaching $y_c=-1$, the evolution of the force shows significant qualitative differences depending on the Froude number (see also Vincent {\em et al.} \cite{lionel}). This behavior is observed for both water and oil. In particular, below $Fr \approx 4$ the force abruptly increases shortly before and during interface crossing, reaching a maximum after the cylinder has exited the bath. This force increase reflects the fact that the reduction of the buoyancy contribution dominates the reduction of the drag when the cylinder exits the bath. For Froude numbers larger than $Fr \approx 4$, the situation is inverted: the decrease of the drag force is larger than that of the buoyancy so that the overall force decreases during the interface crossing. At $Fr \approx 4$ (or $Fr \approx 5$ for the oil), the combined contribution of buoyancy and drag for the cylinder in the liquid bath is more or less equal to the weight of the entrained liquid when the cylinder is out of the bath.  

 Once the cylinder has emerged out of the bath, the measured force corresponds to the weight of the cylinder (minus the buoyancy) plus the weight of the entrained liquid. Consequently, following the increase at low $Fr$ and decrease at large $Fr$, the force converges to similar values for all Froude numbers, but remaining nonetheless larger for higher Froude numbers owing the the larger amount of liquid entrained at larger speeds. During drainage, the force then slowly and monotonously decreases towards the equilibrium value, i.e., the weight of the cylinder minus the buoyancy force (this follows from the definition of the reference force $F=0$~N). The bump observed for $Fr \lesssim 4$ just after interface crossing corresponds thus to the weight of the entrained liquid. Lastly, note that the temporary force decrease at large $y_c$ (dashed lines in Fig.~\ref{fig:force}(b)) is due to inertial effects during the deceleration of the cylinder.

Finally, the force in oil is reported as a function of time in Fig.~\ref{fig:zoom} when the cylinder is out of the liquid bath, namely during the drainage phase, $t>0$~s.  The force measurements $F(t)$ for all Froude numbers seem to converge towards a common curve. The dashed curve represents a power law, i.e., $F \propto t^{-1/2}$. This power law is compatible with the observed law of thickness decay found in Fig.~\ref{fig:whole} and with the common decay law found in the literature (e.g. \cite{drainage}).

\section{Conclusion}\label{conclusion}
In this paper, an experimental setup has been developed to study the forced exit dynamics of a fully submerged horizontal cylinder. Two cases were considered: the water and the silicone oil (50 cSt) for Froude numbers lower than 15.

The aspect ratio of the cylinder was taken as large as possible and equipped with end-plates to observe as close as possible two-dimensional flow conditions at the center of the cylinder. The best combination was checked by analyzing the flow in the liquid bath through PIV measurements and by evaluating the dependence of the interface displacement on the aspect ratio. 

The wake of the cylinder was studied during the displacement of the cylinder in the bath. PIV measurements showed the formation of clockwise and anti-clockwise vortices during the cylinder's upward motion. The vortices shed in the wake were quickly diffused and large vortices were only observed in the vicinity of the cylinder. These vortices remained symmetrically positioned in the case of the silicone oil (lower Reynolds number), and this even for long-traveling cases ($d>20a$). In the case of water, the vortices were detached and asymmetrical.

The thickness of the fluid layer above the cylinder was also measured through two different techniques: image analysis when the thickness was above 1~mm and Chromatic Confocal Point Sensor for smaller thicknesses during the drainage phase. The combination and the synchronization of both data allowed evidence of a change of regime in the thinning of the entrained liquid at the apex of the cylinder. Just after the interface crossing, the thickness of the liquid decreased according to an exponential decay in time. For times larger than $\simeq$ 250~ms and thicknesses lower than $\simeq$ 1~mm, the thinning followed a power scaling law of the form $t^{-1/2}$. 

Finally, the force measurements carried out during the upward motion of the $AR=12$ cylinder allowed the estimation of the net drag force. As expected, results showed that the net drag force and entrained force increased with the cylinder's upward velocity. Consequently, there exists a particular Froude number above which the force drops after the crossing of the interface. The force measurements also allowed us to confirm that the drainage dynamics evolved according to the square root of the time. 

In the future, the overall deformation of the interface due to more complex shaped objects should be studied. Indeed, when the object approaches the interface, the interface first deforms according to a bump before revealing the shape of the object during the drainage stage. The force variations could also indicate the influence of the wake on the liquid entrainment. However, the speed of the crossing has been confirmed to have a minimal impact on the drainage dynamics of the fluid around the object.

\section*{Acknowledgements} The financial support of the Belgian Fund for Scientific Research under research project WOLFLOW (F.R.S.-FNRS, PDR T.0021.18) is gratefully acknowledged. Part of the experimental setup was funded by the {\it Fonds Sp\'eciaux} from the Universit\'e de Li\`{e}ge. SD and BS are  FNRS senior research associates and research directors, respectively.

\bibliography{example}

\begin{thebibliography}{40}%
\makeatletter
\providecommand \@ifxundefined [1]{%
 \@ifx{#1\undefined}
}%
\providecommand \@ifnum [1]{%
 \ifnum #1\expandafter \@firstoftwo
 \else \expandafter \@secondoftwo
 \fi
}%
\providecommand \@ifx [1]{%
 \ifx #1\expandafter \@firstoftwo
 \else \expandafter \@secondoftwo
 \fi
}%
\providecommand \natexlab [1]{#1}%
\providecommand \enquote  [1]{``#1''}%
\providecommand \bibnamefont  [1]{#1}%
\providecommand \bibfnamefont [1]{#1}%
\providecommand \citenamefont [1]{#1}%
\providecommand \href@noop [0]{\@secondoftwo}%
\providecommand \href [0]{\begingroup \@sanitize@url \@href}%
\providecommand \@href[1]{\@@startlink{#1}\@@href}%
\providecommand \@@href[1]{\endgroup#1\@@endlink}%
\providecommand \@sanitize@url [0]{\catcode `\\12\catcode `\$12\catcode `\&12\catcode `\#12\catcode `\^12\catcode `\_12\catcode `\%12\relax}%
\providecommand \@@startlink[1]{}%
\providecommand \@@endlink[0]{}%
\providecommand \url  [0]{\begingroup\@sanitize@url \@url }%
\providecommand \@url [1]{\endgroup\@href {#1}{\urlprefix }}%
\providecommand \urlprefix  [0]{URL }%
\providecommand \Eprint [0]{\href }%
\providecommand \doibase [0]{https://doi.org/}%
\providecommand \selectlanguage [0]{\@gobble}%
\providecommand \bibinfo  [0]{\@secondoftwo}%
\providecommand \bibfield  [0]{\@secondoftwo}%
\providecommand \translation [1]{[#1]}%
\providecommand \BibitemOpen [0]{}%
\providecommand \bibitemStop [0]{}%
\providecommand \bibitemNoStop [0]{.\EOS\space}%
\providecommand \EOS [0]{\spacefactor3000\relax}%
\providecommand \BibitemShut  [1]{\csname bibitem#1\endcsname}%
\let\auto@bib@innerbib\@empty
\bibitem [{\citenamefont {Chang}\ \emph {et~al.}(2019)\citenamefont {Chang}, \citenamefont {Myeong}, \citenamefont {Virot}, \citenamefont {Clanet}, \citenamefont {Kim},\ and\ \citenamefont {Jung}}]{chang2019jumping}%
  \BibitemOpen
  \bibfield  {author} {\bibinfo {author} {\bibfnamefont {B.}~\bibnamefont {Chang}}, \bibinfo {author} {\bibfnamefont {J.}~\bibnamefont {Myeong}}, \bibinfo {author} {\bibfnamefont {E.}~\bibnamefont {Virot}}, \bibinfo {author} {\bibfnamefont {C.}~\bibnamefont {Clanet}}, \bibinfo {author} {\bibfnamefont {H.-Y.}\ \bibnamefont {Kim}},\ and\ \bibinfo {author} {\bibfnamefont {S.}~\bibnamefont {Jung}},\ }\bibfield  {title} {\bibinfo {title} {Jumping dynamics of aquatic animals},\ }\href@noop {} {\bibfield  {journal} {\bibinfo  {journal} {Journal of the Royal Society Interface}\ }\textbf {\bibinfo {volume} {16}},\ \bibinfo {pages} {20190014} (\bibinfo {year} {2019})}\BibitemShut {NoStop}%
\bibitem [{\citenamefont {Chang}\ \emph {et~al.}(2016)\citenamefont {Chang}, \citenamefont {Croson}, \citenamefont {Straker},\ and\ \citenamefont {Jung}}]{dive}%
  \BibitemOpen
  \bibfield  {author} {\bibinfo {author} {\bibfnamefont {B.}~\bibnamefont {Chang}}, \bibinfo {author} {\bibfnamefont {M.}~\bibnamefont {Croson}}, \bibinfo {author} {\bibfnamefont {L.}~\bibnamefont {Straker}},\ and\ \bibinfo {author} {\bibfnamefont {S.}~\bibnamefont {Jung}},\ }\bibfield  {title} {\bibinfo {title} {How seabirds plunge-dive without injuries},\ }\href@noop {} {\bibfield  {journal} {\bibinfo  {journal} {Proceedings of the National Academy of Sciences of the United States of America}\ }\textbf {\bibinfo {volume} {113}},\ \bibinfo {pages} {12006} (\bibinfo {year} {2016})}\BibitemShut {NoStop}%
\bibitem [{\citenamefont {Challa}\ \emph {et~al.}(2014)\citenamefont {Challa}, \citenamefont {Yim}, \citenamefont {Idichandy},\ and\ \citenamefont {Vendhan}}]{challa2014}%
  \BibitemOpen
  \bibfield  {author} {\bibinfo {author} {\bibfnamefont {R.}~\bibnamefont {Challa}}, \bibinfo {author} {\bibfnamefont {S.~C.}\ \bibnamefont {Yim}}, \bibinfo {author} {\bibfnamefont {V.}~\bibnamefont {Idichandy}},\ and\ \bibinfo {author} {\bibfnamefont {C.}~\bibnamefont {Vendhan}},\ }\bibfield  {title} {\bibinfo {title} {Rigid-object water-entry impact dynamics: Finite-element/smoothed particle hydrodynamics modeling and experimental validation},\ }\href@noop {} {\bibfield  {journal} {\bibinfo  {journal} {Journal of Offshore Mechanics and Arctic Engineering}\ }\textbf {\bibinfo {volume} {136}},\ \bibinfo {pages} {031102} (\bibinfo {year} {2014})}\BibitemShut {NoStop}%
\bibitem [{\citenamefont {Challa}\ \emph {et~al.}(2010)\citenamefont {Challa}, \citenamefont {Idichandy}, \citenamefont {Vendhan},\ and\ \citenamefont {Yim}}]{challa2010}%
  \BibitemOpen
  \bibfield  {author} {\bibinfo {author} {\bibfnamefont {R.}~\bibnamefont {Challa}}, \bibinfo {author} {\bibfnamefont {V.}~\bibnamefont {Idichandy}}, \bibinfo {author} {\bibfnamefont {C.}~\bibnamefont {Vendhan}},\ and\ \bibinfo {author} {\bibfnamefont {S.}~\bibnamefont {Yim}},\ }\bibfield  {title} {\bibinfo {title} {An experimental study on rigid-object water-entry impact and contact dynamics},\ }in\ \href@noop {} {\emph {\bibinfo {booktitle} {ASME 2010 29th International Conference on Ocean, Offshore and Arctic Engineering}}}\ (\bibinfo {organization} {American Society of Mechanical Engineers Digital Collection},\ \bibinfo {year} {2010})\ pp.\ \bibinfo {pages} {383--391}\BibitemShut {NoStop}%
\bibitem [{\citenamefont {Mohtat}\ \emph {et~al.}(2015)\citenamefont {Mohtat}, \citenamefont {Challa}, \citenamefont {Yim},\ and\ \citenamefont {Judge}}]{mohtat2015}%
  \BibitemOpen
  \bibfield  {author} {\bibinfo {author} {\bibfnamefont {A.}~\bibnamefont {Mohtat}}, \bibinfo {author} {\bibfnamefont {R.}~\bibnamefont {Challa}}, \bibinfo {author} {\bibfnamefont {S.~C.}\ \bibnamefont {Yim}},\ and\ \bibinfo {author} {\bibfnamefont {C.~Q.}\ \bibnamefont {Judge}},\ }\bibfield  {title} {\bibinfo {title} {Numerical modeling of hydrodynamic impact and local slamming effects},\ }in\ \href@noop {} {\emph {\bibinfo {booktitle} {Proceedings of the 13th International Conference on Fast Sea Transportation, Washington DC, Estados Unidos}}}\ (\bibinfo {year} {2015})\BibitemShut {NoStop}%
\bibitem [{\citenamefont {Yang}\ and\ \citenamefont {Qiu}(2012)}]{yang2012}%
  \BibitemOpen
  \bibfield  {author} {\bibinfo {author} {\bibfnamefont {Q.}~\bibnamefont {Yang}}\ and\ \bibinfo {author} {\bibfnamefont {W.}~\bibnamefont {Qiu}},\ }\bibfield  {title} {\bibinfo {title} {Numerical simulation of water impact for 2d and 3d bodies},\ }\href@noop {} {\bibfield  {journal} {\bibinfo  {journal} {Ocean Engineering}\ }\textbf {\bibinfo {volume} {43}},\ \bibinfo {pages} {82} (\bibinfo {year} {2012})}\BibitemShut {NoStop}%
\bibitem [{\citenamefont {Nair}\ and\ \citenamefont {Bhattacharyya}(2018)}]{nair2018water}%
  \BibitemOpen
  \bibfield  {author} {\bibinfo {author} {\bibfnamefont {V.~V.}\ \bibnamefont {Nair}}\ and\ \bibinfo {author} {\bibfnamefont {S.}~\bibnamefont {Bhattacharyya}},\ }\bibfield  {title} {\bibinfo {title} {Water entry and exit of axisymmetric bodies by cfd approach},\ }\href@noop {} {\bibfield  {journal} {\bibinfo  {journal} {Journal of Ocean Engineering and Science}\ }\textbf {\bibinfo {volume} {3}},\ \bibinfo {pages} {156} (\bibinfo {year} {2018})}\BibitemShut {NoStop}%
\bibitem [{\citenamefont {Zhou}\ \emph {et~al.}(2024)\citenamefont {Zhou}, \citenamefont {Zhao}, \citenamefont {Dai}, \citenamefont {Yao}, \citenamefont {Liu}, \citenamefont {Zhang}, \citenamefont {Wang},\ and\ \citenamefont {Zhang}}]{projectile}%
  \BibitemOpen
  \bibfield  {author} {\bibinfo {author} {\bibfnamefont {B.}~\bibnamefont {Zhou}}, \bibinfo {author} {\bibfnamefont {Z.}~\bibnamefont {Zhao}}, \bibinfo {author} {\bibfnamefont {Q.}~\bibnamefont {Dai}}, \bibinfo {author} {\bibfnamefont {W.}~\bibnamefont {Yao}}, \bibinfo {author} {\bibfnamefont {X.}~\bibnamefont {Liu}}, \bibinfo {author} {\bibfnamefont {Y.}~\bibnamefont {Zhang}}, \bibinfo {author} {\bibfnamefont {A.}~\bibnamefont {Wang}},\ and\ \bibinfo {author} {\bibfnamefont {H.}~\bibnamefont {Zhang}},\ }\bibfield  {title} {\bibinfo {title} {{Numerical study on the cavity dynamics of water entry and exit for a high-speed projectile crossing a wave}},\ }\href@noop {} {\bibfield  {journal} {\bibinfo  {journal} {Physics of Fluids}\ }\textbf {\bibinfo {volume} {36}},\ \bibinfo {pages} {063321} (\bibinfo {year} {2024})}\BibitemShut {NoStop}%
\bibitem [{\citenamefont {Landau}\ and\ \citenamefont {Levich}(1988)}]{LL}%
  \BibitemOpen
  \bibfield  {author} {\bibinfo {author} {\bibfnamefont {L.}~\bibnamefont {Landau}}\ and\ \bibinfo {author} {\bibfnamefont {V.}~\bibnamefont {Levich}},\ }\bibfield  {title} {\bibinfo {title} {Dragging of a liquid by a moving plate},\ }\href@noop {} {\bibfield  {journal} {\bibinfo  {journal} {Dynamics of Curved Fronts}\ ,\ \bibinfo {pages} {141}} (\bibinfo {year} {1988})}\BibitemShut {NoStop}%
\bibitem [{\citenamefont {Landau}\ and\ \citenamefont {Levich}(1942)}]{LL2}%
  \BibitemOpen
  \bibfield  {author} {\bibinfo {author} {\bibfnamefont {L.}~\bibnamefont {Landau}}\ and\ \bibinfo {author} {\bibfnamefont {V.}~\bibnamefont {Levich}},\ }\bibfield  {title} {\bibinfo {title} {Dragging of a liquid by a moving plate},\ }\href@noop {} {\bibfield  {journal} {\bibinfo  {journal} {Acta Physicochim. URSS}\ }\textbf {\bibinfo {volume} {17}},\ \bibinfo {pages} {42} (\bibinfo {year} {1942})}\BibitemShut {NoStop}%
\bibitem [{\citenamefont {de~Ryck}\ and\ \citenamefont {Qu\'er\'e}(1998)}]{deryck}%
  \BibitemOpen
  \bibfield  {author} {\bibinfo {author} {\bibfnamefont {A.}~\bibnamefont {de~Ryck}}\ and\ \bibinfo {author} {\bibfnamefont {D.}~\bibnamefont {Qu\'er\'e}},\ }\bibfield  {title} {\bibinfo {title} {Fluid coating from a polymer solution},\ }\href@noop {} {\bibfield  {journal} {\bibinfo  {journal} {Langmuir}\ }\textbf {\bibinfo {volume} {14}},\ \bibinfo {pages} {1911} (\bibinfo {year} {1998})}\BibitemShut {NoStop}%
\bibitem [{\citenamefont {Seiwert}\ \emph {et~al.}(2011)\citenamefont {Seiwert}, \citenamefont {Clanet},\ and\ \citenamefont {Qu\'er\'e}}]{seiwert}%
  \BibitemOpen
  \bibfield  {author} {\bibinfo {author} {\bibfnamefont {J.}~\bibnamefont {Seiwert}}, \bibinfo {author} {\bibfnamefont {C.}~\bibnamefont {Clanet}},\ and\ \bibinfo {author} {\bibfnamefont {D.}~\bibnamefont {Qu\'er\'e}},\ }\bibfield  {title} {\bibinfo {title} {Coating of a textured solid},\ }\href@noop {} {\bibfield  {journal} {\bibinfo  {journal} {Journal of Fluid Mechanics}\ }\textbf {\bibinfo {volume} {669}},\ \bibinfo {pages} {55} (\bibinfo {year} {2011})}\BibitemShut {NoStop}%
\bibitem [{\citenamefont {Dixit}\ and\ \citenamefont {Homsy}(2013)}]{dixit}%
  \BibitemOpen
  \bibfield  {author} {\bibinfo {author} {\bibfnamefont {H.}~\bibnamefont {Dixit}}\ and\ \bibinfo {author} {\bibfnamefont {G.}~\bibnamefont {Homsy}},\ }\bibfield  {title} {\bibinfo {title} {The elastocapillary landau-levich problem},\ }\href@noop {} {\bibfield  {journal} {\bibinfo  {journal} {Journal of Fluid Mechanics}\ }\textbf {\bibinfo {volume} {735}},\ \bibinfo {pages} {1} (\bibinfo {year} {2013})}\BibitemShut {NoStop}%
\bibitem [{\citenamefont {Benilov}\ \emph {et~al.}(2010)\citenamefont {Benilov}, \citenamefont {Chapman}, \citenamefont {McLeod}, \citenamefont {Ocknedon},\ and\ \citenamefont {V.S.}}]{benilov}%
  \BibitemOpen
  \bibfield  {author} {\bibinfo {author} {\bibfnamefont {E.}~\bibnamefont {Benilov}}, \bibinfo {author} {\bibfnamefont {S.}~\bibnamefont {Chapman}}, \bibinfo {author} {\bibfnamefont {J.}~\bibnamefont {McLeod}}, \bibinfo {author} {\bibfnamefont {J.}~\bibnamefont {Ocknedon}},\ and\ \bibinfo {author} {\bibfnamefont {Z.}~\bibnamefont {V.S.}},\ }\bibfield  {title} {\bibinfo {title} {On liquid films on an inclined},\ }\href@noop {} {\bibfield  {journal} {\bibinfo  {journal} {Journal of Fluid Mechanics}\ }\textbf {\bibinfo {volume} {663}},\ \bibinfo {pages} {53} (\bibinfo {year} {2010})}\BibitemShut {NoStop}%
\bibitem [{\citenamefont {Weinstein}\ and\ \citenamefont {Ruschak}(2001)}]{weinstein}%
  \BibitemOpen
  \bibfield  {author} {\bibinfo {author} {\bibfnamefont {S.}~\bibnamefont {Weinstein}}\ and\ \bibinfo {author} {\bibfnamefont {K.}~\bibnamefont {Ruschak}},\ }\bibfield  {title} {\bibinfo {title} {Dip coating on a planar non-vertical substrate in the limit of negligible surface tension},\ }\href@noop {} {\bibfield  {journal} {\bibinfo  {journal} {Chemical Engineering Science}\ }\textbf {\bibinfo {volume} {56}},\ \bibinfo {pages} {4957} (\bibinfo {year} {2001})}\BibitemShut {NoStop}%
\bibitem [{\citenamefont {Jambon-Puillet}\ \emph {et~al.}(2021)\citenamefont {Jambon-Puillet}, \citenamefont {Ledda}, \citenamefont {Gallaire},\ and\ \citenamefont {Brun}}]{jambon}%
  \BibitemOpen
  \bibfield  {author} {\bibinfo {author} {\bibfnamefont {E.}~\bibnamefont {Jambon-Puillet}}, \bibinfo {author} {\bibfnamefont {P.~G.}\ \bibnamefont {Ledda}}, \bibinfo {author} {\bibfnamefont {F.}~\bibnamefont {Gallaire}},\ and\ \bibinfo {author} {\bibfnamefont {P.}~\bibnamefont {Brun}},\ }\bibfield  {title} {\bibinfo {title} {Drops on the underside of a slightly inclined wet substrate move too fast to grow},\ }\href@noop {} {\bibfield  {journal} {\bibinfo  {journal} {Physical Review Letters}\ }\textbf {\bibinfo {volume} {127}},\ \bibinfo {pages} {044503} (\bibinfo {year} {2021})}\BibitemShut {NoStop}%
\bibitem [{\citenamefont {Eghbali}\ \emph {et~al.}(2024)\citenamefont {Eghbali}, \citenamefont {Djambov},\ and\ \citenamefont {Gallaire}}]{eghbali}%
  \BibitemOpen
  \bibfield  {author} {\bibinfo {author} {\bibfnamefont {S.}~\bibnamefont {Eghbali}}, \bibinfo {author} {\bibfnamefont {S.}~\bibnamefont {Djambov}},\ and\ \bibinfo {author} {\bibfnamefont {F.}~\bibnamefont {Gallaire}},\ }\bibfield  {title} {\bibinfo {title} {Stability of a liquid layer draining around a horizontal cylinder: Interplay of capillary and gravity forces},\ }\href@noop {} {\bibfield  {journal} {\bibinfo  {journal} {Physical Review Fluids}\ }\textbf {\bibinfo {volume} {9}},\ \bibinfo {pages} {063903} (\bibinfo {year} {2024})}\BibitemShut {NoStop}%
\bibitem [{\citenamefont {Reynolds}(1886)}]{Reynolds}%
  \BibitemOpen
  \bibfield  {author} {\bibinfo {author} {\bibfnamefont {O.}~\bibnamefont {Reynolds}},\ }\bibfield  {title} {\bibinfo {title} {On the theory of lubrication and its application to mr beauchamp tower's experiments, including an experimental determination of the viscsoity of olive oil},\ }\href@noop {} {\bibfield  {journal} {\bibinfo  {journal} {Philosophical Transactions of the Royal Society of London}\ }\textbf {\bibinfo {volume} {AI 77}},\ \bibinfo {pages} {157} (\bibinfo {year} {1886})}\BibitemShut {NoStop}%
\bibitem [{\citenamefont {Chan}\ and\ \citenamefont {Horn}(1985)}]{drainage}%
  \BibitemOpen
  \bibfield  {author} {\bibinfo {author} {\bibfnamefont {D.}~\bibnamefont {Chan}}\ and\ \bibinfo {author} {\bibfnamefont {R.}~\bibnamefont {Horn}},\ }\bibfield  {title} {\bibinfo {title} {The drainage of thin liquid films tween solid surfaces},\ }\href@noop {} {\bibfield  {journal} {\bibinfo  {journal} {The Journal of Chemical Physics}\ }\textbf {\bibinfo {volume} {83}},\ \bibinfo {pages} {5311} (\bibinfo {year} {1985})}\BibitemShut {NoStop}%
\bibitem [{\citenamefont {Coons}\ \emph {et~al.}(2003)\citenamefont {Coons}, \citenamefont {Halley}, \citenamefont {McGlashan},\ and\ \citenamefont {Tran-Cong}}]{coons}%
  \BibitemOpen
  \bibfield  {author} {\bibinfo {author} {\bibfnamefont {J.}~\bibnamefont {Coons}}, \bibinfo {author} {\bibfnamefont {P.}~\bibnamefont {Halley}}, \bibinfo {author} {\bibfnamefont {S.}~\bibnamefont {McGlashan}},\ and\ \bibinfo {author} {\bibfnamefont {T.}~\bibnamefont {Tran-Cong}},\ }\bibfield  {title} {\bibinfo {title} {A review of drainage and spontaneous rupture in free standing thin films with tangentially immobile interfaces},\ }\href@noop {} {\bibfield  {journal} {\bibinfo  {journal} {Advances in Colloid and Interface Science}\ }\textbf {\bibinfo {volume} {105}},\ \bibinfo {pages} {3} (\bibinfo {year} {2003})}\BibitemShut {NoStop}%
\bibitem [{\citenamefont {Bhamla}\ \emph {et~al.}(2014)\citenamefont {Bhamla}, \citenamefont {Giacomin}, \citenamefont {Balemans},\ and\ \citenamefont {Fuller}}]{bhamla}%
  \BibitemOpen
  \bibfield  {author} {\bibinfo {author} {\bibfnamefont {M.}~\bibnamefont {Bhamla}}, \bibinfo {author} {\bibfnamefont {C.}~\bibnamefont {Giacomin}}, \bibinfo {author} {\bibfnamefont {C.}~\bibnamefont {Balemans}},\ and\ \bibinfo {author} {\bibfnamefont {G.}~\bibnamefont {Fuller}},\ }\bibfield  {title} {\bibinfo {title} {Influence of interfacial rheology on drainage from curved surfaces},\ }\href@noop {} {\bibfield  {journal} {\bibinfo  {journal} {Soft Matter}\ }\textbf {\bibinfo {volume} {10}},\ \bibinfo {pages} {6917} (\bibinfo {year} {2014})}\BibitemShut {NoStop}%
\bibitem [{\citenamefont {Wu}\ \emph {et~al.}(2017)\citenamefont {Wu}, \citenamefont {Ni}, \citenamefont {Bai}, \citenamefont {Cui},\ and\ \citenamefont {Sun}}]{wu2017experimental}%
  \BibitemOpen
  \bibfield  {author} {\bibinfo {author} {\bibfnamefont {Q.}~\bibnamefont {Wu}}, \bibinfo {author} {\bibfnamefont {B.}~\bibnamefont {Ni}}, \bibinfo {author} {\bibfnamefont {X.}~\bibnamefont {Bai}}, \bibinfo {author} {\bibfnamefont {B.}~\bibnamefont {Cui}},\ and\ \bibinfo {author} {\bibfnamefont {S.}~\bibnamefont {Sun}},\ }\bibfield  {title} {\bibinfo {title} {Experimental study on large deformation of free surface during water exit of a sphere},\ }\href@noop {} {\bibfield  {journal} {\bibinfo  {journal} {Ocean Engineering}\ }\textbf {\bibinfo {volume} {140}},\ \bibinfo {pages} {369} (\bibinfo {year} {2017})}\BibitemShut {NoStop}%
\bibitem [{\citenamefont {Liju}\ \emph {et~al.}(2001)\citenamefont {Liju}, \citenamefont {Machane},\ and\ \citenamefont {Cartellier}}]{liju2001}%
  \BibitemOpen
  \bibfield  {author} {\bibinfo {author} {\bibfnamefont {P.-Y.}\ \bibnamefont {Liju}}, \bibinfo {author} {\bibfnamefont {R.}~\bibnamefont {Machane}},\ and\ \bibinfo {author} {\bibfnamefont {A.}~\bibnamefont {Cartellier}},\ }\bibfield  {title} {\bibinfo {title} {Surge effect during the water exit of an axisymmetric body traveling normal to a plane interface: experiments and bem simulation},\ }\href@noop {} {\bibfield  {journal} {\bibinfo  {journal} {Experiments in fluids}\ }\textbf {\bibinfo {volume} {31}},\ \bibinfo {pages} {241} (\bibinfo {year} {2001})}\BibitemShut {NoStop}%
\bibitem [{\citenamefont {Moshari}\ \emph {et~al.}(2014)\citenamefont {Moshari}, \citenamefont {Nikseresht},\ and\ \citenamefont {Mehryar}}]{moshari2014}%
  \BibitemOpen
  \bibfield  {author} {\bibinfo {author} {\bibfnamefont {S.}~\bibnamefont {Moshari}}, \bibinfo {author} {\bibfnamefont {A.~H.}\ \bibnamefont {Nikseresht}},\ and\ \bibinfo {author} {\bibfnamefont {R.}~\bibnamefont {Mehryar}},\ }\bibfield  {title} {\bibinfo {title} {Numerical analysis of two and three dimensional buoyancy driven water-exit of a circular cylinder},\ }\href@noop {} {\bibfield  {journal} {\bibinfo  {journal} {International Journal of Naval Architecture and Ocean Engineering}\ }\textbf {\bibinfo {volume} {6}},\ \bibinfo {pages} {219} (\bibinfo {year} {2014})}\BibitemShut {NoStop}%
\bibitem [{\citenamefont {Truscott}\ \emph {et~al.}(2016)\citenamefont {Truscott}, \citenamefont {Epps},\ and\ \citenamefont {Munns}}]{truscott2016}%
  \BibitemOpen
  \bibfield  {author} {\bibinfo {author} {\bibfnamefont {T.~T.}\ \bibnamefont {Truscott}}, \bibinfo {author} {\bibfnamefont {B.~P.}\ \bibnamefont {Epps}},\ and\ \bibinfo {author} {\bibfnamefont {R.~H.}\ \bibnamefont {Munns}},\ }\bibfield  {title} {\bibinfo {title} {Water exit dynamics of buoyant spheres},\ }\href@noop {} {\bibfield  {journal} {\bibinfo  {journal} {Physical Review Fluids}\ }\textbf {\bibinfo {volume} {1}},\ \bibinfo {pages} {074501} (\bibinfo {year} {2016})}\BibitemShut {NoStop}%
\bibitem [{\citenamefont {Bourrier}\ \emph {et~al.}(1984)\citenamefont {Bourrier}, \citenamefont {Guyon},\ and\ \citenamefont {Jorre}}]{guyon}%
  \BibitemOpen
  \bibfield  {author} {\bibinfo {author} {\bibfnamefont {P.}~\bibnamefont {Bourrier}}, \bibinfo {author} {\bibfnamefont {E.}~\bibnamefont {Guyon}},\ and\ \bibinfo {author} {\bibfnamefont {J.}~\bibnamefont {Jorre}},\ }\bibfield  {title} {\bibinfo {title} {The pop off effect: different regimes of a light ball in water},\ }\href@noop {} {\bibfield  {journal} {\bibinfo  {journal} {European Journal of Physics}\ }\textbf {\bibinfo {volume} {5.4}},\ \bibinfo {pages} {225} (\bibinfo {year} {1984})}\BibitemShut {NoStop}%
\bibitem [{\citenamefont {Takamure}\ and\ \citenamefont {Uchiyama}(2021)}]{takamure2021effect}%
  \BibitemOpen
  \bibfield  {author} {\bibinfo {author} {\bibfnamefont {K.}~\bibnamefont {Takamure}}\ and\ \bibinfo {author} {\bibfnamefont {T.}~\bibnamefont {Uchiyama}},\ }\bibfield  {title} {\bibinfo {title} {Effect of froude number on the motion of a spherical particle launched vertically upward in water},\ }\href@noop {} {\bibfield  {journal} {\bibinfo  {journal} {Experimental Thermal and Fluid Science}\ }\textbf {\bibinfo {volume} {128}},\ \bibinfo {pages} {110453} (\bibinfo {year} {2021})}\BibitemShut {NoStop}%
\bibitem [{\citenamefont {Ashraf}\ and\ \citenamefont {Dorbolo}(2024{\natexlab{a}})}]{intesaaf1}%
  \BibitemOpen
  \bibfield  {author} {\bibinfo {author} {\bibfnamefont {I.}~\bibnamefont {Ashraf}}\ and\ \bibinfo {author} {\bibfnamefont {S.}~\bibnamefont {Dorbolo}},\ }\bibfield  {title} {\bibinfo {title} {Effect of the surface dimples on the exit dynamics of a sphere at a constant velocity},\ }\href@noop {} {\bibfield  {journal} {\bibinfo  {journal} {Applied Ocean Research}\ }\textbf {\bibinfo {volume} {147}},\ \bibinfo {pages} {103996} (\bibinfo {year} {2024}{\natexlab{a}})}\BibitemShut {NoStop}%
\bibitem [{\citenamefont {Chu}\ \emph {et~al.}(2010)\citenamefont {Chu}, \citenamefont {Yan}, \citenamefont {Wang}, \citenamefont {Zhang}, \citenamefont {Feng},\ and\ \citenamefont {Chen}}]{chu2010}%
  \BibitemOpen
  \bibfield  {author} {\bibinfo {author} {\bibfnamefont {X.-s.}\ \bibnamefont {Chu}}, \bibinfo {author} {\bibfnamefont {K.}~\bibnamefont {Yan}}, \bibinfo {author} {\bibfnamefont {Z.}~\bibnamefont {Wang}}, \bibinfo {author} {\bibfnamefont {K.}~\bibnamefont {Zhang}}, \bibinfo {author} {\bibfnamefont {G.}~\bibnamefont {Feng}},\ and\ \bibinfo {author} {\bibfnamefont {W.-q.}\ \bibnamefont {Chen}},\ }\bibfield  {title} {\bibinfo {title} {Numerical simulation of water-exit of a cylinder with cavities},\ }\href@noop {} {\bibfield  {journal} {\bibinfo  {journal} {Journal of Hydrodynamics, Ser. B}\ }\textbf {\bibinfo {volume} {22}},\ \bibinfo {pages} {877} (\bibinfo {year} {2010})}\BibitemShut {NoStop}%
\bibitem [{\citenamefont {Havelock}(1936)}]{Havelock1936}%
  \BibitemOpen
  \bibfield  {author} {\bibinfo {author} {\bibfnamefont {T.~H.}\ \bibnamefont {Havelock}},\ }\bibfield  {title} {\bibinfo {title} {The forces on a circular cylinder submerged in a uniform stream},\ }\href@noop {} {\bibfield  {journal} {\bibinfo  {journal} {Proceedings of the royal society A}\ } (\bibinfo {year} {1936})}\BibitemShut {NoStop}%
\bibitem [{\citenamefont {Greenhow}\ and\ \citenamefont {Lin}(1983)}]{greenhow1983nonlinear}%
  \BibitemOpen
  \bibfield  {author} {\bibinfo {author} {\bibfnamefont {M.}~\bibnamefont {Greenhow}}\ and\ \bibinfo {author} {\bibfnamefont {W.-M.}\ \bibnamefont {Lin}},\ }\href@noop {} {\emph {\bibinfo {title} {Nonlinear-free surface effects: experiments and theory}}},\ \bibinfo {type} {Tech. Rep.}\ (\bibinfo  {institution} {Massachusetts Inst Of Tech Cambridge Dept Of Ocean Engineering},\ \bibinfo {year} {1983})\BibitemShut {NoStop}%
\bibitem [{\citenamefont {Greenhow}\ and\ \citenamefont {Moyo}(1997)}]{greenhow1997}%
  \BibitemOpen
  \bibfield  {author} {\bibinfo {author} {\bibfnamefont {M.}~\bibnamefont {Greenhow}}\ and\ \bibinfo {author} {\bibfnamefont {S.}~\bibnamefont {Moyo}},\ }\bibfield  {title} {\bibinfo {title} {Water entry and exit of horizontal circular cylinders},\ }\href@noop {} {\bibfield  {journal} {\bibinfo  {journal} {Philosophical Transactions of the Royal Society of London. Series A: Mathematical, Physical and Engineering Sciences}\ }\textbf {\bibinfo {volume} {355}},\ \bibinfo {pages} {551} (\bibinfo {year} {1997})}\BibitemShut {NoStop}%
\bibitem [{\citenamefont {Telste}(1987)}]{telste1987}%
  \BibitemOpen
  \bibfield  {author} {\bibinfo {author} {\bibfnamefont {J.~G.}\ \bibnamefont {Telste}},\ }\bibfield  {title} {\bibinfo {title} {Inviscid flow about a cylinder rising to a free surface},\ }\href@noop {} {\bibfield  {journal} {\bibinfo  {journal} {Journal of Fluid Mechanics}\ }\textbf {\bibinfo {volume} {182}},\ \bibinfo {pages} {149} (\bibinfo {year} {1987})}\BibitemShut {NoStop}%
\bibitem [{\citenamefont {Ni}\ \emph {et~al.}(2015)\citenamefont {Ni}, \citenamefont {Zhang},\ and\ \citenamefont {Wu}}]{ni2015}%
  \BibitemOpen
  \bibfield  {author} {\bibinfo {author} {\bibfnamefont {B.}~\bibnamefont {Ni}}, \bibinfo {author} {\bibfnamefont {A.}~\bibnamefont {Zhang}},\ and\ \bibinfo {author} {\bibfnamefont {G.}~\bibnamefont {Wu}},\ }\bibfield  {title} {\bibinfo {title} {Simulation of complete water exit of a fully-submerged body},\ }\href@noop {} {\bibfield  {journal} {\bibinfo  {journal} {Journal of Fluids and Structures}\ }\textbf {\bibinfo {volume} {58}},\ \bibinfo {pages} {79} (\bibinfo {year} {2015})}\BibitemShut {NoStop}%
\bibitem [{\citenamefont {Ashraf}\ and\ \citenamefont {Dorbolo}(2024{\natexlab{b}})}]{intesaaf2}%
  \BibitemOpen
  \bibfield  {author} {\bibinfo {author} {\bibfnamefont {I.}~\bibnamefont {Ashraf}}\ and\ \bibinfo {author} {\bibfnamefont {S.}~\bibnamefont {Dorbolo}},\ }\bibfield  {title} {\bibinfo {title} {Exit dynamics of a square cylinder},\ }\href@noop {} {\bibfield  {journal} {\bibinfo  {journal} {Ocean Engineering}\ }\textbf {\bibinfo {volume} {297}},\ \bibinfo {pages} {117106} (\bibinfo {year} {2024}{\natexlab{b}})}\BibitemShut {NoStop}%
\bibitem [{\citenamefont {Haohao}\ \emph {et~al.}(2019)\citenamefont {Haohao}, \citenamefont {Yanping}, \citenamefont {Jianyang}, \citenamefont {Fu},\ and\ \citenamefont {Tian}}]{haohao2019numerical}%
  \BibitemOpen
  \bibfield  {author} {\bibinfo {author} {\bibfnamefont {H.}~\bibnamefont {Haohao}}, \bibinfo {author} {\bibfnamefont {S.}~\bibnamefont {Yanping}}, \bibinfo {author} {\bibfnamefont {Y.}~\bibnamefont {Jianyang}}, \bibinfo {author} {\bibfnamefont {C.}~\bibnamefont {Fu}},\ and\ \bibinfo {author} {\bibfnamefont {L.}~\bibnamefont {Tian}},\ }\bibfield  {title} {\bibinfo {title} {Numerical analysis of water exit for a sphere with constant velocity using the lattice boltzmann method},\ }\href@noop {} {\bibfield  {journal} {\bibinfo  {journal} {Applied Ocean Research}\ }\textbf {\bibinfo {volume} {84}},\ \bibinfo {pages} {163} (\bibinfo {year} {2019})}\BibitemShut {NoStop}%
\bibitem [{\citenamefont {Miao}(1989)}]{miao1989}%
  \BibitemOpen
  \bibfield  {author} {\bibinfo {author} {\bibfnamefont {G.}~\bibnamefont {Miao}},\ }\bibfield  {title} {\bibinfo {title} {Hydrodynamic forces and dynamic responses of circular cylinders in wave zones},\ }\href@noop {} {\bibfield  {journal} {\bibinfo  {journal} {Ph.D. thesis, Dept. of Marine Hydrodynamics, NTH, Trondheim.}\ } (\bibinfo {year} {1989})}\BibitemShut {NoStop}%
\bibitem [{\citenamefont {Wei}\ \emph {et~al.}(2024)\citenamefont {Wei}, \citenamefont {Li}, \citenamefont {Lei}, \citenamefont {Li}, \citenamefont {Riviero-Rodriguez}, \citenamefont {Lin}, \citenamefont {Wang},\ and\ \citenamefont {Scheid}}]{wei2024}%
  \BibitemOpen
  \bibfield  {author} {\bibinfo {author} {\bibfnamefont {X.}~\bibnamefont {Wei}}, \bibinfo {author} {\bibfnamefont {D.}~\bibnamefont {Li}}, \bibinfo {author} {\bibfnamefont {J.}~\bibnamefont {Lei}}, \bibinfo {author} {\bibfnamefont {J.}~\bibnamefont {Li}}, \bibinfo {author} {\bibfnamefont {J.}~\bibnamefont {Riviero-Rodriguez}}, \bibinfo {author} {\bibfnamefont {F.}~\bibnamefont {Lin}}, \bibinfo {author} {\bibfnamefont {D.}~\bibnamefont {Wang}},\ and\ \bibinfo {author} {\bibfnamefont {B.}~\bibnamefont {Scheid}},\ }\bibfield  {title} {\bibinfo {title} {Exit dynamics of a sphere launched underneath a liquid bath surface},\ }\href@noop {} {\bibfield  {journal} {\bibinfo  {journal} {Physical Review Fluids}\ }\textbf {\bibinfo {volume} {9}},\ \bibinfo {pages} {054003} (\bibinfo {year} {2024})}\BibitemShut {NoStop}%
\bibitem [{\citenamefont {Thielicke}\ and\ \citenamefont {Stamhuis}(2014)}]{thielicke2014pivlab}%
  \BibitemOpen
  \bibfield  {author} {\bibinfo {author} {\bibfnamefont {W.}~\bibnamefont {Thielicke}}\ and\ \bibinfo {author} {\bibfnamefont {E.}~\bibnamefont {Stamhuis}},\ }\bibfield  {title} {\bibinfo {title} {Pivlab--towards user-friendly, affordable and accurate digital particle image velocimetry in matlab},\ }\href@noop {} {\bibfield  {journal} {\bibinfo  {journal} {Journal of open research software}\ }\textbf {\bibinfo {volume} {2}} (\bibinfo {year} {2014})}\BibitemShut {NoStop}%
\bibitem [{\citenamefont {Vincent}\ \emph {et~al.}(tted)\citenamefont {Vincent}, \citenamefont {Rivero}, \citenamefont {Falla}, \citenamefont {Ashraf}, \citenamefont {Terrapon}, \citenamefont {Dorbolo},\ and\ \citenamefont {Scheid}}]{lionel}%
  \BibitemOpen
  \bibfield  {author} {\bibinfo {author} {\bibfnamefont {L.}~\bibnamefont {Vincent}}, \bibinfo {author} {\bibfnamefont {J.}~\bibnamefont {Rivero}}, \bibinfo {author} {\bibfnamefont {R.}~\bibnamefont {Falla}}, \bibinfo {author} {\bibfnamefont {I.}~\bibnamefont {Ashraf}}, \bibinfo {author} {\bibfnamefont {V.~E.}\ \bibnamefont {Terrapon}}, \bibinfo {author} {\bibfnamefont {S.}~\bibnamefont {Dorbolo}},\ and\ \bibinfo {author} {\bibfnamefont {B.}~\bibnamefont {Scheid}},\ }\bibfield  {title} {\bibinfo {title} {Inertial exit dynamics of a horizontal cylinder out of a liquid bath},\ }\href@noop {} {\bibfield  {journal} {\bibinfo  {journal} {Journal of Fluid Mechanics}\ } (\bibinfo {year} {submitted})}\BibitemShut {NoStop}%
\end{thebibliography}%

\end{document}